\documentclass[12pt]{article}

\newcommand{\blind}{0}

\usepackage[utf8]{inputenc}
\usepackage{comment}

\usepackage{lscape} 
\usepackage{varioref}
\usepackage{appendix}
\usepackage{natbib}
\usepackage{multirow}
\usepackage{booktabs}
\usepackage[flushleft]{threeparttable}
\usepackage{amsmath}
\usepackage{amssymb}
\usepackage{amsthm}
\usepackage{graphicx}
\usepackage{float}
\usepackage{subfig}
\usepackage{soul}
\usepackage{xcolor}
\usepackage[linesnumbered,ruled,vlined]{algorithm2e}
\usepackage[linktocpage=true]{hyperref}

\usepackage[capitalise,noabbrev]{cleveref}

\theoremstyle{plain}

\theoremstyle{definition}

\newcommand{\vect}[1]{{\boldsymbol{#1}}} 
\newcommand{\indep}{\perp \!\!\! \perp} 
\newcommand{\lo}{\text{low}} 
\newcommand{\hi}{\text{high}} 
\def\one{\mbox{1\hspace{-4.25pt}\fontsize{12}{14.4}\selectfont\textrm{1}}} 
\def\spacingset#1{\renewcommand{\baselinestretch}%
{#1}\small\normalsize} \spacingset{1}

\begin{document}

\title{Individualized Prediction Bands in Causal Inference with Continuous Treatments}

\if0\blind
{
  \title{\bf Individualized Prediction Bands in Causal Inference with Continuous Treatments }
  \author{Max Sampson \footnote{Max Sampson was supported by National Institutes of Health Predoctoral Training Grant T32 HL 144461.
}\\
    Department of Statistics, Florida State University\\
    and \\
    Kung-Sik Chan \\
    Department of Statistics and Actuarial Science, University of Iowa}
  \maketitle
} \fi

\if1\blind
{
  \bigskip
  \bigskip
  \bigskip
  \begin{center}
    {\LARGE\bf Title}
\end{center}
  \medskip
} \fi

\bigskip

\begin{abstract}
Individualized treatments are crucial for optimal decision-making and treatment allocation, specifically in personalized medicine based on the estimation of an individual's dose-response curve across a continuum of treatment levels, e.g., drug dosage. Current works focus on conditional mean and median estimates, which are useful but do not provide the full picture. We propose viewing causal inference with a continuous treatment as a covariate shift. This allows us to leverage existing weighted conformal prediction methods with both quantile and point estimates to compute individualized uncertainty quantification for dose-response curves. When used with quantile regression, our method, individualized prediction bands (IPB), enjoys a doubly robust coverage property. We demonstrate IPB via simulations and a real data analysis that assesses the additional medical expenditure caused by continued smoking for selected individuals. Our results show that IPB provides an effective solution to a gap in the individual dose-response uncertainty quantification literature. 
\end{abstract}


\section{Introduction}

Throughout this paper  we focus on the potential outcome framework \citep{neyman1923, rubin1978} with a continuous treatment \citep{imbens2000, HiranoandImbens2004}. With \(n\) subjects, denote the continuous treatment by \(T_i \in \mathcal{T} = [t_{\min}, t_{\max}]\), \(\mathcal{Y}_i = \{Y_i(t) : t \in \mathcal{T}\}\), the potential outcome at treatment $t$, and $\vect{X}_i$, a vector of $p$ covariates. Note the potential outcome for the ith individual at treatment level $t$, $Y_i(t)$, is different than the observed response, $Y_i$. $Y_i(t)$ is what the response for the ith individual would have been if their treatment was $t$ and not $T_i$. 

For example, let $T$ be the time from the onset of hearing loss to a cochlear implant, $X$ is the gender (which we assume is the only variable that directly affects the treatment and the response), and $Y$ is a measure of hearing function. It is of interest to find the time window for a cochlear implant to ``recover'' a certain level of hearing function. That is, what is a realistic value of time after hearing loss, $t$, so that $Y(t)$ is as large as possible? As another example, let $T$ be the average yearly smoking load, $\vect{X}$ demographic and socioeconomic variables, and $Y$ the subject's medical costs over a certain period. The problem is to assess how much of the medical costs were directly caused by smoking versus not smoking, $Y_i - Y_i(0)$. These questions are important on a population level: what is the optimal average time window for a cochlear implant? They are also very important on an individual level: what is the optimal time window for this specific patient to receive a cochlear implant? These questions can be answered with counterfactual dose-response functions, which we break up into population level average dose-response functions,
\[
\mu(t) = E(Y(t)),
\]
conditional average dose-response functions,
\[
\mu(t\mid  \vect{X}) = E(Y(t) \mid  \vect{X}),
\]
and individual dose-response functions,
\[
Y_i(t) \mid \vect{X}_i.
\]
In this paper we focus on the latter, individual dose-response functions. To our knowledge, there are two existing methods that combine conformal prediction \citep{split_conformal_lei_2016, conformal_book, romanocqr} and causal inference with a continuous treatment. The first, \cite{schroder2025conformal}, focuses on creating prediction intervals for causal effects, e.g., $Y(t^*) - Y(t)$, and not the dose-response curve as we do. The second, \cite{verhaeghe2024conformal}, is more comparable to our approach because they create prediction intervals and bands for the dose-response curve. However, an advantage of our proposed method is that it is doubly robust, see Proposition 2 below; see Section~\ref{sec:comparison with early works} for further delineation of the differences between their approach and ours.  In the remainder of the introduction, we introduce the notational and ideological framework of our problem, as well as useful background information. In~\cref{sec:methods} we discuss existing methods that form causal prediction intervals and bands for individuals when the treatment is continuous, as well as formally introduce our method, Individualized Prediction Bands (IPB). We then demonstrate the usefulness of IPB via numerical experiments and a real application which looks at the economic benefits quitting smoking can have in~\cref{sec:empircal_performance}. 

\subsection{Causal Inference Assumptions}

For a causal statement to be made using dose-response functions, as in the case with a binary/discrete treatment, we need certain assumptions. Following  \cite{rosenbaum_rubin_1983, rosenbaum_rubin1984}, \cite{Lechner_2001}, and \cite{ite_schwab}, we assume unconfoundedness, positivity, and the stable unit treatment value assumption (SUTVA). Strong unconfoundedness is the assumption that the treatment is independent of all potential outcomes conditional on $\vect{X}$. That is, any relationship between potential outcomes and the treatment can be explained by measured covariates. The assumption written out is
\[
\mathcal{Y}\indep T\mid  \vect{X}.
\]
Strong unconfoundedness implies that all confounders have been measured and are present in $\vect{X}$. Weak unconfoundedness is sometimes assumed instead, for example in \cite{imbens2000, HiranoandImbens2004}. Weak unconfoundedness is the assumption that the treatment is independent of each potential outcome conditional on $\vect{X}$, instead of independent of all potential outcomes. 

Positivity, or the common support assumption, is the assumption that it must be possible to observe all possible treatment levels with a probability greater than zero. With a discrete number of treatments positivity is
\[
0 < P(T = t\mid \vect{X}) < 1 \text{ a.s., }  \forall  t \in \{t_1, \ldots, t_m\}.
\]
In the continuous treatment case, positivity is
\[
P(T \in \mathcal{A} \mid \vect{X}) > 0 \text{ a.s, for every measurable set } \mathcal{A} \in \mathcal{T}  \text{with positive Lebesgue measure}
 \]
 \citep{imaidyke2004}. Positivity can be further broken up into two types: deterministic positivity (that it is possible for any individual to receive any treatment) and random positivity (that in the observed sample, individuals with similar covariates have received multiple values of a treatment) \citep{westreich_random_positivity_2010}.

The assumption of SUTVA has two parts. The first part is no interference. That is, the treatment assigned to one individual will not affect the outcome of another individual. The second part is consistency. Consistency requires the treatment to be well defined, so there is only one version of each treatment value. For example, taking a new drug to reduce fevers means the pills contain exactly the same dosage at each dose level \citep{rosenbaum_rubin_1983, rosenbaum_rubin1984}. SUTVA implies that the observed response can be written as the counterfactual of the observed treatment, $Y_i = Y_i(T_i)$. These assumptions will be implicitly made throughout the paper. 

\subsection{Traditional Inferential Targets}

The majority of existing methods in causal inference with continuous treatments focus on the average dose-response curve for a population or a subset of the population. There has been work done on average dose-response functions using parametric, semi-parametric, and non-parametric methods \citep{imbens2000, HiranoandImbens2004, imaidyke2004, bart, flores2012, diaz2013targeted, galvao2015uniformly, kennedy2017non, ZhaoVanDykImai2020, hudson2024approach}, as well as works that have studied the conditional and individual dose-response functions \citep{imbens2009, shalit2017, ite_schwab}.

The methods that investigate individual dose-response functions lack reliable uncertainty estimates. One solution to this for general regression problems with a continuous response has been to use quantile regression \citep{koenker1978}. This has been used to estimate the conditional quantile treatment effect (CQTE) in causal inference \citep{fort2016}. In the binary case, the CQTE is defined as the difference of the $\beta$-th quantiles between $Y(1)$ and $Y(0)$ conditioned on $\vect{X} = \vect{x}$. Though, reliable uncertainty estimates are still needed for CQTE as well as quantile regression done with continuous treatments.

Motivated by the question of: Given a possible range of treatments, what are reasonably plausible potential outcomes for these treatments? We consider the problem of estimating conditional quantiles of individual dose-response functions. We propose to do this in two ways, both of which only require observational data. The first constructs prediction bands about an average dose-response function utilizing weighted conformal prediction. The second uses weighted conformal prediction to adjust conditional quantiles, forming prediction bands. To form these individualized prediction bands, we view observational causal inference with a continuous treatment as a covariate shift.

\subsubsection{Interval-Based Decision Rules}

If the objective is to  identify a set of treatments that will result in the potential outcome exceeding a certain threshold, interval-based decision rules can be used \citep{chen2022IBDR}. \cite{park2025personalizedtwosideddoseinterval} significantly extended this framework to hormetic dose-response relationships, which follow an inverted U-shaped curve. Their approach finds the treatment bounds simultaneously, instead of calculating the lower and upper bounds of the intervals iteratively, as in \cite{chen2022IBDR}. 

To find a range of treatments that contains the treatment that maximizes the potential outcome, the method introduced by \cite{lu2023_jump} can be used. This approach breaks the treatment's range into disjoint intervals. Within each of these intervals, the maximum potential outcome is found. This results in a treatment interval which contains the highest estimated potential outcome for an individual.

\subsection{Counterfactuals and Covariate Shift}\label{sec: counterfactual and covariate shift}

One way to view counterfactual inference is as a question: ``What if the treatment was applied according to a distribution independent of confounders?'' This question is important to keep in mind when constructing estimates for $Y(t), \text{ }t \in \mathcal{T}$ using the observed i.i.d. observations, $(Y_i^{obs}, T_i)$. Consider a simple example of how sugar consumption affects heart disease. A person who consumes a lot of sugar may be less likely to exercise regularly than someone who is health conscious and does not consume a lot of sugar. If we ignore the amount someone exercises, our causal estimate of sugar consumption on heart health will be biased. If we control for how often a person exercises, along with all other variables that affect both how often a person consumes sugar and their heart health, we would be able to estimate the causal effect of sugar on heart disease, and how changes in sugar consumption can causally affect heart disease. 

To conduct counterfactual inference, we can view the observed distribution as
\[
(Y_i, T_i, \vect{X}_i) \overset{i.i.d.}{\sim} P_{\vect{X}} \times P_{T\mid \vect{X}} \times P_{Y\mid \vect{X},T}, \quad i = 1, 2, \ldots, n,
\]
where $P_\cdot$ and $P_{\cdot\mid \cdot}$ denote the marginal probability distribution and the conditional probability distribution, respectively. The treatment being assigned independently of confounders, as is the case in randomized experiments, can be viewed as a covariate shift from our original distribution:
\[
(Y_i, T_i, \vect{X}_i)\overset{i.i.d.}{\sim} P_{\vect{X}} \times Q_{T} \times P_{Y\mid \vect{X},T}, \quad i = n+1, n+2, \ldots ,
\]
where $Q_T$ is the new distribution of $T$ after the shift has occurred, for instance, the observed marginal distribution of $T$, and the conditional distribution, $P_{Y\mid  \vect{X}, T}$, and marginal distribution, $P_{\vect{X}}$, remain the same. For example, what would the effect on hearing be if we gave someone a cochlear implant quickly after hearing loss began, as opposed to several years later, assuming all else is equal?

This will allow us to infer what happens when the treatment is assigned independently of any covariates. Point estimates for covariate shifts have been widely studied in machine learning \citep{SHIMODAIRA2000_covariate_shift}. Interval estimates (both confidence and prediction intervals) estimates are, however, understudied in the literature. Weighted conformal inference has recently allowed prediction intervals to be created from data with known covariate shifts \citep{conformal_shift} and unknown, but estimated, covariate shifts \citep{cqr, qin2025distribution}. 

A special case of $Q_T$, when the focus is on a potential outcome at a single specific treatment $t$, is the degenerate distribution represented by the Dirac delta function, $\delta_t$. However, applying this in practice with a continuous treatment, in both randomized trials and observational studies, would limit the effective sample size because the weights for every observation where $T_i \neq t$ would be zero. Furthermore, as noted by \citet{schindl2026incremental}, specifying a single treatment in the counterfactual world does not always provide useful information for practitioners. 

Another scenario where specifying $Q_T$ is useful is when a randomized trial cannot be conducted, but is of interest. For example, with large observational datasets, $Q_T$ is specified as the random treatment assignment distribution \citep{hernan2016using}. In a similar vein, because $Q_T$ is practitioner-specified, this framework can be extended to methods with stochastic interventions, including incremental propensity score interventions with both categorical and continuous treatments \citep{munoz2012population, kennedy2019nonparametric, schindl2026incremental}. 

This weighting strategy naturally extends to randomized and conditionally randomized trials where a continuous dosage is discretized. However, this discretization can lead to poor inference if the dose-response curve lacks sufficient smoothness and monotonicity. Specifically, smoothness is required to rule out step-wise functions, which cannot be interpolated easily between the grid points, and monotonicity ensures inflection points (for the optimal treatment) are not obscured by the grid. 

We combine the idea of viewing causal inference with a continuous treatment as a covariate shift with conformal prediction to form individual prediction bands to answer the question, ``What if the treatment was assigned independently of the confounders?''

\subsection{Conformal Prediction}

Conformal prediction is a general method of creating prediction intervals that provide a non-asymptotic, distribution-free coverage guarantee. We focus on using split conformal prediction to form prediction bands.

Split conformal prediction has three steps. The first is to split the data into a training set, $Z_{tr}$, and a calibration set, $Z_{cal}$. The second step trains a model on the training set. The model chosen depends on what non-conformity score one chooses to work with. For a continuous response, conditional mean and quantile regression models are common choices \citep{split_conformal_lei_2016, conformal_book, romanocqr}. The third step computes the non-conformity scores on the calibration data. A common non-conformity score for a regression problem is the absolute difference, $V_i = |Y_i - \hat{f}(\vect{X}_i)|$ \citep{papadopoulos_2002}. The prediction interval that is output will depend on the non-conformity score used; in the case of the absolute difference, the interval is
\[
C(\vect{X}_{n+1}) = [\hat{f}(\vect{X}_{n+1}) - Q_{1-\alpha}(\vect{V}; Z_{cal}), \hat{f}(\vect{X}_{n+1}) + Q_{1-\alpha}(\vect{V}; Z_{cal})],
\]
where 
\[
Q_{1-\alpha}(\vect{V}; Z_{cal}) := (1-\alpha)(1 + \frac{1}{|Z_{cal}|})-\text{th empirical quantile of }\{V_i\},
\]
and $|Z_{cal}|$ is the size of the calibration set. 
The process of split conformal prediction described above is formalized in \Cref{alg:conformal_example}.

\begin{algorithm}[H]
\caption{Split Conformal Prediction}\label{alg:conformal_example}
\KwIn{Significance level $\alpha$, dataset $\mathcal{Z} = \{(Y_i, \vect{X}_i)\}_{i \in \mathcal{I}}$, test point $\vect{x}$, and regression algorithm $f(\vect{X}; \mathcal{D})$}
\KwOut{Prediction interval $\hat{C}(\vect{x})$}

\SetKwProg{Fn}{Procedure}{}{}
\Fn{}{
    Split $\mathcal{Z}$ into a training set $\mathcal{Z}_{tr} = \{(Y_i, \vect{X}_i)\}_{i \in \mathcal{I}_{tr}}$ and a calibration set $\mathcal{Z}_{cal} = \{(Y_i, \vect{X}_i)\}_{i \in \mathcal{I}_{cal}}$\;
    
    Train the model $\hat{f}(\vect{x}; \mathcal{Z}_{tr})$ using the training set\;
    
    \ForEach{$i \in \mathcal{I}_{cal}$}{
        Compute score $V_i = |\hat{f}(\vect{X}_i; \mathcal{Z}_{tr}) - Y_i|$\;
    }
    
    Compute $\eta(\vect{x})$ as the $(1 - \alpha)$-quantile of the scores $\{V_i\}$\;
}
\Return{$\hat{C}(\vect{x}) = [\hat{f}(\vect{x}; \mathcal{Z}_{tr}) - \eta(\vect{x}), \hat{f}(\vect{x}; \mathcal{Z}_{tr}) + \eta(\vect{x})]$}
\end{algorithm}

Conformal prediction can also be used with conditional quantile regression \citep{romanocqr, cqr}. The steps are the same as with a point estimate, except the trained model is used to estimate lower and upper conditional quantiles, $\hat{q}_{\alpha_{low}}$ and $\hat{q}_{\alpha_{high}}$. The non-conformity scores are then, $V_i = \text{max}\{\hat{q}_{\alpha_{low}}(\vect{X}_i) - Y_i, Y_i - \hat{q}_{\alpha_{high}}(\vect{X}_i)\}$. Finally, for the new input, $\vect{X}_{n+1}$, the prediction interval for $Y_{n+1}$ is 
\[
C(\vect{X}_{n+1}) = [\hat{q}_{\alpha_{low}}(\vect{X}_{n+1}) - Q_{1-\alpha}(\vect{V}; Z_{cal}),\hat{q}_{\alpha_{high}} + Q_{1-\alpha}(\vect{V}; Z_{cal})]
\]
where 
\[
Q_{1-\alpha}(\vect{V}; Z_{cal}) := (1-\alpha)(1 + \frac{1}{|Z_{cal}|})-\text{th empirical quantile of }\{V_i\}.
\]

To form valid prediction bands, conditional coverage over all possible treatments should be guaranteed. However, achieving finite sample, distribution-free, conditional coverage for a continuous response,
\[
P(Y_{n+1} \in C(\vect{X}_{n+1})\mid \vect{X}_{n+1}) \geq 1-\alpha, \quad a.s.
\]
results in prediction intervals with an infinite expected length \citep{dis_free_pred_COPS}. We address this by utilizing conformalized quantile regression with IPB, which leverages the doubly robust properties established in \cite{cqr}. These properties ensure a coverage guarantee when either the conditional distribution of the treatment, conditional quantile outcome regression model, or both are correctly specified. Furthermore, conformalized quantile regression (CQR) provides a conditional coverage guarantee when the conditional quantile model is correctly specified. We also provide a method to form individualized prediction bands that starts with a point estimate. The main idea of the next section is that the non-conformity scores can be shifted to form prediction intervals when covariate shifts occur. So, by viewing the causal inference problem as a covariate shift, we can leverage weighted conformal prediction to form individual prediction bands.

\subsection{Weighted Conformal Inference} \label{sec:weighted_conformal}

Our goal is to shift the treatment distribution from being drawn from the conditional distribution of $T\mid \vect{X}$ to being drawn from an assignment distribution that does not depend on confounders, for example the marginal treatment distribution $f_T$. To do this, following \cite{conformal_shift, cqr}, we need to look at the likelihood ratio of the shifted distributions to find the weight. Similar to importance sampling, the distribution of interest is in the numerator and the observed distribution is in the denominator.

\begin{equation}\label{eq:importance_weight}
\frac{dQ_{T, \vect{X}}}{dP_{T, \vect{X}}} = \frac{f_{\vect{X}}f_T }{f_\vect{X} f_{T\mid \vect{X}}} = \frac{f_T}{f_{T\mid \vect{X}}}
\end{equation}

For a continuous treatment, this is the same as the stabilized weights introduced in \cite{RobbinsHernan2000MSM}. This ratio is the same as in the setting of a binary treatment in \cite{cqr}, though in our setting with a continuous treatment, we are unable to group treatments together to ignore the marginal distribution like we can in the binary case \citep{cqr}. Instead of directly using the marginal distribution in the numerator of the weight, we use an assignment distribution. The assignment or intervention distribution is the distribution by which the treatment is assigned. This provides the flexibility to ask questions such as, ``what if we truncate possible values of the treatment'' or ``what if we don't randomly assign the treatment, but assign it conditionally on observed covariates in a specific way''. The conditional distribution of a continuous treatment is known as the generalized propensity score (GPS), and is commonly used in causal inference with continuous treatments. A more detailed discussion of the GPS and its use in causal inference with continuous treatments can be found in Appendix D of the supplementary materials. Our focus is on combining the GPS with an outcome regression model to form individualized prediction bands. For a discussion of estimating the GPS as well as the role the GPS plays in the bias of causal methods see \cite{tu2019comparison, huling2024independence}. 

\subsection{Outcome Regression Models}

Outcome regression models are commonly used to model the conditional mean of the response, $Y$, given the variables, $\vect{X}$, that are necessary to satisfy the strong ignorability assumption along with the treatment, $T$ \citep{cibook}. For example, if the treatment and covariates are linearly related to the response, one could use ordinary least squares to estimate $\beta_0,$ $\beta_1, $ and $\beta_2$ in the equation below.
\[
E(Y\mid T, X) = \beta_0 + \beta_1T + \beta_2X
\]
If the model is correctly specified, it can be used to make causal claims, assuming that all causal assumptions (e.g., unconfoundedness, positivity, etc.) are true. This is due to the fact that the assumption of consistency implies that the conditional expectation of a potential outcome at a given treatment level is the same as the observed response given the treatment level. That is, $E(Y(t)\mid T=t,\vect{X}) = E(Y\mid T=t, \vect{X})$. \cite{bart} used this approach with Bayesian Additive Regression Trees to look at population level comparisons between treatments.

This idea can be extended beyond the conditional mean if we use a stronger form of consistency. Using this consistency assumption with strong ignorability, we can show that a properly specified outcome regression model can tell us about the potential outcome,
\begin{align*}
    & Y(t)\mid \vect{X} \\
    & \overset{d}{=} Y(t)\mid \vect{X}, T = t \\ 
    & \overset{d}{=} Y\mid \vect{X}, T = t,
\end{align*}
where the first equality is by the strong ignorability assumption and the second equality is by the stronger consistency assumption. This brief explanation tells us that if we are able to model the conditional distribution of $Y\mid \vect{X}, T$ correctly, we can use it to make valid causal claims. 

We are not the first to make the argument that outcome regression models can be extended beyond the conditional mean. \cite{Zhang_Chen_2012} argue that for population causal effects, such as the average treatment effect, the median may be a better measure than the mean for heavily skewed distributions. This can be generalized to quantiles other than the median. For example, \cite{fort2016} looked at the conditional quantiles to estimate conditional quantile treatment effects. 

We build on these approaches by combining conditional quantile regression with weighted conformal prediction in IPB. Combining conditional quantile regression with weighted conformal prediction provides us robustness against misspecified conditional quantile models, and has a negligible impact when the conditional model is correctly specified. 

\section{Methods} \label{sec:methods}

\subsection{Individual Prediction Bands with Continuous Treatments}\label{sec: IPB}
 
We outline our method, Individual Prediction Bands with Continuous Treatments (IPB), below. IPB combines the outcome regression model for a continuous treatment with the generalized propensity score for individual causal predictions to form individual doubly robust prediction intervals, as well as individual dose-response prediction bands. When IPB is used with a good conditional quantile model, the resulting prediction bands have near $1-\alpha$ coverage over all treatments of interest. When IPB is paired with a point estimation (such as a conditional mean regression model), IPB loses the conditional coverage guarantee, but has a marginal coverage guarantee and interpretation, which is what we refer to as the doubly robust property. Both of these guarantees are explained in more details after our introduction of IPB, which is given below. 

On its own, looking at a prediction interval  for a treatment drawn from the marginal distribution may not be of much interest. We are more likely to be interested in what happens if we assign a specific treatment, or treatments in a small interval near each other. Instead of letting $dQ_{T, \vect{X}} = f_{\vect{X}}f_T$, let $dQ_{T, \vect{X}} = f_{\vect{X}}h_T$ where $h_T$ has no reliance on $\vect{X}$ and has a more concentrated density near some point, $t$, than $f_T$. Generally, $h_T$ is known and is henceforth known as the intervention or treatment assignment distribution. To motivate this idea, consider the following example.

Imagine a person tears their ACL and their doctor wants to know what the optimal waiting time is for their surgery. They decide that the best time to give the surgery is two weeks after the injury has occurred so the individual is able to let the swelling reduce, a common approach for ACL tears \citep{shen_liu_ACL_2022}. Because of the surgeon's schedule, the surgery cannot always be scheduled exactly two weeks from the injury. There also may be times when the swelling does not go down, and the surgery needs to be postponed for a few days \citep{shen_liu_ACL_2022}. So, we would like to create a prediction set that gives a high likelihood of receiving the surgery around two weeks after injury, but that has positive probability for that individual to receive the surgery at a later time because of unforeseen complications. This  idea is further explored numerically in~\cref{sec:treat_trunc}.

Over all treatment values, we are interested in creating a prediction interval with coverage guarantees. Our observed sample data do not match the data we wish to predict because we are not observing a natural treatment based on a set of covariates. This was addressed by \cite{conformal_shift} by introducing a weighted version on conformal inference with a known likelihood ratio, 
\[w(\cdot) = \frac{dQ}{dP} (\cdot).\]
Herein and henceforth, $\frac{dQ}{dP}$ is given by \eqref{eq:importance_weight} with $f_T$ replaced by $h_T$. This idea was further extended in \cite{cqr} to the scenario where the likelihood ratio was unknown. They specifically looked at a binary treatment with quantile regression modeling the response. We extend this to a continuous treatment with any split conformal method, as well as specifically with split conformal quantile regression. 

 ~\cref{alg:mean_reg} sketches the method for any split conformal method. ~\cref{alg:quant_reg} shows the method as applied to split conformal quantile regression, which we use in the majority of our numerical studies. We note that these methods can also use a signed-conformal regression approach to form one-sided prediction intervals \citep{Linusson_2014_signed_conformal, romanocqr}. These algorithms can be viewed as weighting the scores in conformal prediction, similar to importance weighting of the response done with other causal inference methods. 

The weights we use are,
\[
W_i = \hat{w}(\vect{X}_i, T_i; Z_{tr}) = \frac{h_T(T_i)}{\hat{f}_{T \mid \vect{X}}(T_i \mid \vect{X}_i)} \quad \forall i\in \mathcal{I}_{cal},
\]
where $\hat{\cdot}$ represents  a quantity estimated on the training set. 

\begin{algorithm}[t]
\caption{Weighted Conformal Point Regression}\label{alg:mean_reg}
\KwIn{Significance level $\alpha$, dataset $\mathcal{Z} = \{(Y_i, T_i, \vect{X}_i)\}_{i \in \mathcal{I}}$, test point $(\vect{x}, t)$, regression algorithm $f(\vect{X}, t; \mathcal{D})$, weight function $\hat{w}(\vect{x}, t; \mathcal{D})$}
\KwOut{Prediction interval $\hat{C}(\vect{x}, t)$}

\SetKwProg{Fn}{Procedure}{}{}
\Fn{}{
    Split $\mathcal{Z}$ into a training set $\mathcal{Z}_{tr} = \{(Y_i, T_i, \vect{X}_i)\}_{i \in \mathcal{I}_{tr}}$ and a calibration set $\mathcal{Z}_{cal} = \{(Y_i, T_i, \vect{X}_i)\}_{i \in \mathcal{I}_{cal}}$\;
    
    Train the model $\hat{f}(\vect{x}, t; \mathcal{Z}_{tr})$ using the training set\;
    
    \ForEach{$i \in \mathcal{I}_{cal}$}{
        Compute score $V_i = |\hat{f}(\vect{X}_i, T_i; \mathcal{Z}_{tr}) - Y_i|$\;
        Compute weight $W_i = \hat{w}(\vect{X}_i, T_i; \mathcal{Z}_{tr})$\;
    }
    
    Normalize weights:\;
    $\hat{p}_i(\vect{x}, t) = \dfrac{W_i}{\sum_{j \in \mathcal{I}_{cal}} W_j + \hat{w}(\vect{x}, t; \mathcal{Z}_{tr})}$\;
    $\hat{p}_{\infty}(\vect{x}, t) = \dfrac{\hat{w}(\vect{x}, t; \mathcal{Z}_{tr})}{\sum_{j \in \mathcal{I}_{cal}} W_j + \hat{w}(\vect{x}, t; \mathcal{Z}_{tr})}$\;
    
    Compute $\eta(\vect{x}, t)$ as the $(1 - \alpha)$-quantile of the distribution $\sum\limits_{i \in \mathcal{I}_{cal}} \hat{p}_i(\vect{x}, t) \delta_{V_i} + \hat{p}_{\infty}(\vect{x}, t)\delta_{\infty}$, where $\delta$ is the Dirac delta function\;
}
\Return{$\hat{C}(\vect{x}, t) = [\hat{f}(\vect{x}, t; \mathcal{Z}_{tr}) - \eta(\vect{x}, t), \hat{f}(\vect{x}, t; \mathcal{Z}_{tr}) + \eta(\vect{x}, t)]$}
\end{algorithm}

\begin{algorithm}[H]
\caption{Weighted Conformal Quantile Regression}\label{alg:quant_reg}
\KwIn{Significance level $\alpha$, dataset $\mathcal{Z} = \{(Y_i, T_i, \vect{X}_i)\}_{i \in \mathcal{I}}$, test point $(\vect{x}, t)$, quantile regression algorithm $q_{\beta}(\vect{X}, t; \mathcal{D})$, weight function $\hat{w}(\vect{x}, t; \mathcal{D})$}
\KwOut{Prediction interval $\hat{C}(\vect{x}, t)$}

\SetKwProg{Fn}{Procedure}{}{}
\Fn{}{
    Split $\mathcal{Z}$ into a training set $\mathcal{Z}_{tr} = \{(Y_i, T_i, \vect{X}_i)\}_{i \in \mathcal{I}_{tr}}$ and a calibration set $\mathcal{Z}_{cal} = \{(Y_i, T_i, \vect{X}_i)\}_{i \in \mathcal{I}_{cal}}$\;
    
    Train the quantile models $\hat{q}_{\alpha_{low}}(\vect{x}, t; \mathcal{Z}_{tr})$ and $\hat{q}_{\alpha_{high}}(\vect{x}, t; \mathcal{Z}_{tr})$\;
    
    \ForEach{$i \in \mathcal{I}_{cal}$}{
        Compute score $V_i = \max\left\{\hat{q}_{\alpha_{low}}(\vect{X}_i, T_i; \mathcal{Z}_{tr}) - Y_i,\; Y_i - \hat{q}_{\alpha_{high}}(\vect{X}_i, T_i; \mathcal{Z}_{tr})\right\}$\;
        Compute weight $W_i = \hat{w}(\vect{X}_i, T_i; \mathcal{Z}_{tr})$\;
    }
    
    Normalize weights:\;
    $\hat{p}_i(\vect{x}, t) = \dfrac{W_i}{\sum_{j \in \mathcal{I}_{cal}} W_j + \hat{w}(\vect{x}, t; \mathcal{Z}_{tr})}$\;
    $\hat{p}_{\infty}(\vect{x}, t) = \dfrac{\hat{w}(\vect{x}, t; \mathcal{Z}_{tr})}{\sum_{j \in \mathcal{I}_{cal}} W_j + \hat{w}(\vect{x}, t; \mathcal{Z}_{tr})}$\;
    
    Compute $\eta(\vect{x}, t)$ as the $(1 - \alpha)$-quantile of the distribution $\sum\limits_{i \in \mathcal{I}_{cal}} \hat{p}_i(\vect{x}, t) \delta_{V_i} + \hat{p}_{\infty}(\vect{x}, t)\delta_{\infty}$\;
}
\Return{$\hat{C}(\vect{x}, t) = [\hat{q}_{\alpha_{low}}(\vect{x}, t; \mathcal{Z}_{tr}) - \eta(\vect{x}, t),\; \hat{q}_{\alpha_{high}}(\vect{x}, t; \mathcal{Z}_{tr}) + \eta(\vect{x}, t)]$}
\end{algorithm}
If the weights are known, marginal coverage for a population with a treatment that has been assigned according to $Q_T$ is guaranteed. Below we state the theoretical guarantees of these algorithms, which are direct results from \cite{cqr}. ~\cref{alg:continuous} describes how to create a prediction band for the dose-response curve. This can be used for continuous treatments, as well as multi-class treatments.

Proposition 1. \emph{Consider~\cref{alg:quant_reg}. Assume $(\vect{X}_i, T_i, Y_i) \overset{iid}{\sim} P_{T\mid \vect{X}} \times P_\vect{X} \times P_{Y\mid \vect{X}, T}$} and $w(\cdot)= \frac{dQ_T}{dP_{T\mid \vect{X}}}(\cdot)$.

\begin{enumerate}
    \item If $\hat{w}(\cdot) = w(\cdot)$, then
\[
    \mathbb{P}_{(Y, T, \vect{X})\sim P_{\vect{X}} \times Q_T \times P_{Y\mid \vect{X}, T}}(Y(T) \in \hat{C}(\vect{X}, T)) \geq 1 - \alpha.
\]
    \item If $\hat{w}(\cdot) = w(\cdot)$, the non-conformity scores $\{V_i: i \in \mathcal{I}_{cal}\}$ have no ties almost surely, $Q_T$ is absolutely continuous with respect to $P_{T\mid \vect{X}}$, and $(\mathbb{E}_{\vect{X}, T \sim P_{\vect{X}} \times P_{T\mid \vect{X}}}[\hat{w}(X, T)^r])^{1/r} \leq M_r < \infty$ for some constant $M_r$, then
    \[1 - \alpha \leq  \mathbb{P}_{(Y, T, \vect{X})\sim P_{\vect{X}} \times Q_T \times P_{Y\mid \vect{X}, T}}(Y(T) \in \hat{C}(\vect{X}, T)) \leq 1-\alpha + cn^{1/r - 1},\]
    where $c$ is a positive constant that only depends on $M_r$ and $r$. 
\end{enumerate}

In the general case where $\hat{w}(\cdot) \neq w(\cdot)$, set $\Delta_w = \frac{1}{2} \mathbb{E}_{\vect{X}, T \sim P_{\vect{X}} \times P_{T\mid \vect{X}}}|\hat{w}(\vect{X}, T) - w(\vect{X}, T)|$. Then, coverage is always lower bounded by $1 - \alpha - \Delta_w$ and upper bounded by $1 - \alpha + \Delta_w + cn^{1/r - 1}$ under the same assumptions as in 2. Again, this proposition is directly from \citet[Proposition 1]{cqr}. 

Proposition 1 does not provide a guarantee on the conditional coverage for every $t \in \mathcal{T}$ of the prediction bands. Instead, it provides a coverage guarantee averaged over the distribution of $Q_T$. While this is not the primary goal of a prediction band, it serves as a fallback guarantee if Proposition 2 fails to hold. We can describe the ``doubly robust" guarantee in the setting of the ACL example. Assume that $Q_T$ and its corresponding probability density $h_T$ are selected so that the empirical distribution of treatment assignments matches $Q_T$. In this scenario, the majority of the density mass lies near a 2-week window, with some mass falling slightly outside this window due to clinician and patient scheduling conflicts. In this setting, Proposition 1 ensures that regardless of the conditional quantile model's accuracy, the prediction bands have $(1-\alpha)\times 100\%$ coverage over the distribution of treatments. While Proposition 1 lacks a conditional guarantee, it provides IPB with an interpretable and robust policy-level coverage guarantee when $Q_T$ is carefully selected. 

The next result is directly from \citet[Corollary 1]{cqr}. 

Proposition 2. Denote $N$ as the size of the training set and $n$ as the size of the calibration set.

Assume that either A1 or A2 (or both) is satisfied:\\
\textbf{A1}:
\[
\lim_{N \to \infty} \mathbb{E}|\hat{w}(\vect{X}, T) - w(\vect{X}, T)| = 0,
\]
\textbf{A2:}
\begin{enumerate}
    \item[(1)] $\alpha_{\hi} - \alpha_{\lo} = 1 - \alpha$;

    \item[(2)] there exist $r, b_1, b_2 > 0$ such that the conditional probability density function $f_{Y,\mid \vect{X}, T}(Y = y \mid \vect{X} = \vect{x}, T = t)  \in [b_1, b_2]$ uniformly over all $(\vect{x}, t, y)$ with $y \in [q_{\alpha_{\lo}}(\vect{x}, t) - r, q_{\alpha_{\lo}}(\vect{x}, t) + r] \cup [q_{\alpha_{\hi}}(\vect{x}, t) - r, q_{\alpha_{\hi}}(\vect{x}, t) + r]$;
    \item[(3)] $\mathbb{P}_{\vect{X}, T \sim P_{\vect{X}} \times P_{T\mid \vect{X}}}(w(X) < \infty) = 1$, and there exist $\delta, M > 0$ such that \\$(\mathbb{E}[\hat{w}(\vect{X}, T)^{1+\delta}])^{1/(1+\delta)} \le M$;

    \item[(4)] there exist $k, \ell > 0$ such that\\ $\lim_{N\to\infty} \mathbb{E}[\hat{w}(\vect{X}, T) H_N^k(\vect{X},T)] = \lim_{N\to\infty} \mathbb{E}[w(\vect{X}, T) H_N^\ell(\vect{X}, T)] = 0$, where
    $$
    H_N(\vect{x}, t) = \max\{|\hat{q}_{\alpha_{\lo},N}(\vect{x},t) - q_{\alpha_{\lo}}(\vect{x},t)|, |\hat{q}_{\alpha_{\hi},N}(\vect{x},t) - q_{\alpha_{\hi}}(\vect{x},t)|\}.
    $$
\end{enumerate}
Then, 
\[
\lim_{N, n \to \infty} \mathbb{P}_{(Y, T, \vect{X})\sim P_{\vect{X}} \times Q_T \times P_{Y\mid \vect{X}, T}}(Y(T) \in \hat{C}(\vect{X}, T)) \geq 1 - \alpha.
\]

Furthermore, under \textbf{A2}, for any $\epsilon > 0$,
\[
\lim_{N, n\to \infty} \mathbb{P}_{T, \vect{X} \sim P_{\vect{X}} \times Q_{T}} (\mathbb{P}_{Y\sim P_{Y\mid \vect{X}, T}}(Y(T) \in \hat{C}(\vect{X}, T) \mid \vect{X}, T) \leq 1 - \alpha - \epsilon) = 0.
\]

Proposition 1 implies that with well specified weights, individualized prediction intervals have a marginal coverage over $Q_T$ that is very close to $1 - \alpha$. When the weights are not well specified, the difference between the true coverage and the nominal coverage is bounded by $\Delta_w$. We note that, although not stated, this also applies to the methods of \cite{verhaeghe2024conformal}. Proposition 2 provides the same marginal coverage guarantee and interpretation, and further establishes the much stronger guarantee, which is provided when weighted conformal prediction is combined with quantile regression, that the intervals formed by IPB achieve asymptotic conditional coverage when the conditional quantile model is asymptotically correct. In practice, this means that if a well-specified conditional quantile model is used along with weighted conformal prediction, then the resulting individual prediction band will have conditional coverage over all possible treatments, hence a valid $1-\alpha$ individualized prediction band. 

\begin{algorithm}[H]
\caption{Individualed Prediction Band}\label{alg:continuous}
\KwIn{Significance level $\alpha$, dataset $\mathcal{Z} = \{(Y_i, T_i, \vect{X}_i)\}_{i \in \mathcal{I}}$, test point $\vect{x}$, regression or quantile algorithm for conditional mean or quantile, weight function $\hat{w}(\vect{x}, t; \mathcal{D})$, treatment interval $\mathcal{T} = [t_{\min}, t_{\max}]$, number of splits $N$}
\KwOut{Prediction band $\hat{C}(\vect{x}, \mathcal{T})$}

\SetKwProg{Fn}{Procedure}{}{}
\Fn{}{
    Split $\mathcal{T}$ into $N$ treatment values: $t_1 = t_{\min}, t_2, \ldots, t_N = t_{\max}$\;
    
    \For{$k \gets 1$ \KwTo $N$}{
        Apply either Algorithm~\ref{alg:mean_reg} (Weighted Conformal Point Regression) or Algorithm~\ref{alg:quant_reg} (Weighted Conformal Quantile Regression) at treatment $t_k$\;
        Record the prediction interval $\hat{C}(\vect{x}, t_k)$\;
    }
    
    Combine all prediction intervals $\{\hat{C}(\vect{x}, t_k)\}_{k=1}^N$ into a prediction band $\hat{C}(\vect{x}, \mathcal{T})$\;
}
\Return{$\hat{C}(\vect{x}, \mathcal{T})$}
\end{algorithm}

These algorithms are extensions of the methods discussed in \cite{conformal_shift} with a known likelihood ratio, $w(\cdot)$, and of methods discussed in \cite{cqr} in the case of an unknown, but estimated likelihood ratio, $\hat{w}(\cdot)$. In our case, there are scenarios in which the observed conditional treatment assignment is known, and cases where it needs to be estimated. In both cases, IPB leverages weighted conformal prediction to provide guaranteed coverage prediction bands for individual dose-response curves. These propositions also demonstrate the power of the conformal adjustment: that it improves the coverage when an outcome regression model is misspecified, while maintaining the coverage when the outcome regression model is correctly specified. 

\subsection{Comparison with Existing Methods}~\label{sec:comparison with early works} 

Recall there are two existing methods that combine conformal prediction and causal inference with a continuous treatment:  \cite{schroder2025conformal} and \cite{verhaeghe2024conformal}. We now outline the key differences between the latter approach and ours  below. First, our method can utilize a conditional quantile model as the outcome model, giving our dose-response prediction intervals an asymptotic doubly robust property, that the coverage is guaranteed when either the outcome model or the generalized propensity model is correctly specified. 
Second, they specify two treatment intervention assignments. The first they introduce is a uniform distribution over the treatment values. We note that this can result in highly variable prediction band sizes when the weights are not capped, and it fails to put more mass near the treatment of interest. This is consistent with the  results seen when computing treatment effects from observational data, unstabilized weights (those which use 1 in the numerator instead of the marginal distribution) lead to estimates with higher variability when the weights are not capped \citep{imai2015robust}. 

Their main suggestion is a modified version of the kernel density weight introduced in \cite{conformal_shift} to have better treatment conditional coverage,
\begin{equation}\label{eq:ver_kernel}
w(\vect{X}_i, T_i) = \frac{\one{(T_i \in [t_{\min}, t_{\max}]) \times K(\frac{T_i-t}{h})}}{f_{T\mid \vect{X}}(T_i \mid \vect{X}_i)},
\end{equation}
where $K$ is a kernel density and $h$ its bandwidth. This is done to attempt to improve the conditional coverage for a given treatment. One problem with this approach is that the kernel density estimates need to be computed and stored for each value of $t$ a practitioner plans to investigate.  It is also not clear how to choose the bandwidth $h$ to ensure that the coverage is conditionally valid around the treatment of interest, $t$. We discuss one possible approach in~\cref{sec:treatment assignment}.

We note that this approach is a special case of IPB, where the kernel density estimator serves as the treatment assignment model. We believe that if we are only interested in a single treatment, an approach that is as good, if not better, in this situation is using IPB with a known distribution that puts a majority of its mass near the treatment of interest, $t$. This approach is a more general version of the kernel density approach. For example, a truncated kernel distribution could be used if only a subset of the range of possible treatments is of interest. A skewed distribution could be used if the treatment of interest is near the boundary of treatment values. 

The kernel treatment assignment distribution used by \cite{verhaeghe2024conformal}, given in~\cref{eq:ver_kernel}, has an indicator to ensure the distribution only has positive weight over the range of possible treatment values. This idea can easily be extended to truncate the distribution to a more restricted range of interest. We discuss more weighting options and their implications in~\cref{sec:treatment assignment}. 

We compare four approaches numerically in Appendix B of the supplementary materials, the true treatment assignment distribution, the uniform approach, the kernel approach in~\eqref{eq:ver_kernel}, and a method we introduce in~\cref{sec:treatment assignment}. We find that the approaches yield similar marginal coverage, but the approaches differ in how much weight they assign to the treatment of interest. This demonstrates the importance of determining the question of interest before deciding on the treatment assignment distribution. 

A bimodal distribution specification may also be of interest in certain time-to-treatment (TTT) scenarios. For example, \citet{kaslow2023time} found that the survival rate of those with gastric cancer had a bimodal relationship with TTT. Similarly, for those with acute cholecystitis, \citet{gao2024effect} noted that the optimal timing for emergency surgery is within 72 hours of symptom onset. For patients who miss this early window and instead receive gallbladder drainage (PTGBD) while waiting for surgery, retrospective analysis found that those who delayed the subsequent surgery had better clinical outcomes and patient satisfaction \citep{gao2024effect}. 

\subsubsection{Picking the Treatment Assignment Distribution}\label{sec:treatment assignment}

If the goal is to have conditional coverage over all values of $t \in \mathcal{T}$, and we do not trust our conditional quantile model, the conditional coverage needs to come from the weights. In an ideal world, we would have repeated samples at every possible $t \in \mathcal{T}$, so that we could use a degenerate distribution, $\delta_t$. Because this is not always possible, and is an inefficient use of the data, we  discuss a data driven approach to selecting the variance when the treatment assignment model, $Q_T$ is Normal. This data-driven approach can immediately be extended to selecting the bandwidth of kernel density treatment assignment models. It ensures the treatment assignment distribution places mass from the treatment assignment distribution near the treatment of interest, while maintaining reasonably informative prediction intervals. 

Assume $Q_T$ is $N(t, h)$, where $t$ is the treatment of interest and $h$ is the variance of the distribution. The $N(\cdot, \cdot)$ specification just one shape, it could easily be replaced by a kernel density where $h$ in that case would be the bandwidth instead of the variance, and any number of kernels could be used.  

For conditional coverage at $t$, we want to have as much density near $t$ as possible, which corresponds to having the smallest $h$ possible. Ideally, we would use a degenerate probability distribution to place all the mass at exactly $t$, $\delta_t$. If we do this, we are unable to calibrate our non-conformity scores because no observed data can be reweighted to match this distribution (unless, for some individual $i$, $T_i = t$). Instead, we try to find the smallest variance, $h$, possible.

We can compute the variance, $h$, directly using the training data. This is similar to the leave-one-out approach of \cite{fannjiang2022conformal} but because we already assume a split of the data into a training and a calibration set, we do not need to utilize the leave-one-out approach. 

We find $h$ by solving the following:
\[
\min_{h} \{h:\text{Quantile}(\sum_{i\in \mathcal{I}_{tr}}\hat{p}_i(\vect{x}, t) \delta_{V_i} + \hat{p}_{\infty}(\vect{x}, t)\delta_{\infty}, \> 1-\alpha)<\infty\},
\]
 where $\delta$ is the Dirac delta function; note that $\hat{p}$'s depend on $h$ implicitly. The $1-\alpha$ weighted empirical quantile is $\infty$ only when $\sum_{i\in \mathcal{I}_{tr}}\hat{p}_i(\vect{x}, t) > 1-\alpha$. Thus, for arbitrary but fixed $t$, we can solve the following:
 \[
 \min_{h} \{h: \sum_{i\in \mathcal{I}_{tr}}\hat{p}_i(\vect{x}, t) <1-\alpha\}.
 \]

 If a global $h$ is desired for every treatment of interest, one can grid over the potential treatments $\{t_1, t_2, \ldots, t_N\}$, and for the $k$th treatment, solve the following:
 \[
  h_k=\min_{h} \{h: \sum_{i\in \mathcal{I}_{tr}}\hat{p}_i(\vect{x}, t_k) <1-\alpha\},
 \]
and set the optimal global $h$-value to be  
 \[
 h = \max \{h_1,\ldots, h_N \}.
 \]

 In both cases, especially for small datasets, we would suggest increasing $h$ by a slight percentage (e.g., multiplying it by 1.1) to ensure that the weights are not dominated by a single (or a small number of) observations in the calibration set. 

While not doubly robust in the traditional sense, choosing $h$ in this way allows us to have some marginal interpretation through Proposition 1. If we are unconcerned with this marginal interpretation, we can use $h_k$ with each $t_k$. Using this approach aims to put as much mass near $t$ as possible without having infinitely wide prediction bands. 

There are other weighted conformal approaches that attempt to control conditional coverage near points of interest. Localized conformal prediction, introduced by \cite{guan2023localized} and extended into the randomized setting by \cite{hore2025conformal}, uses a localizer function $H(\vect{x}, \vect{x}')$ that measures the dissimilarity between two features. To be used for causal inference with continuous treatments, one would use a localizer that measures the dissimilarity between two treatment levels, $H(t, t')$. \cite{guan2023localized} showed that using weights which are proportional to a Normal kernel centered at $t$, $\exp(\{T_i-t)^2/h\}$, provides reasonable local coverage. As these weights are not exactly proportional to $\frac{dQ_T}{dP_{T|\mid\vect{X}}}$, the marginal coverage guarantee and interpretation is broken. A modified approach to finding the weighted quantile, compared to that which was introduced in~\cref{sec: IPB}, to provide the marginal coverage and interpretation is provided in \cite{guan2023localized}. 

The randomized approach developed by \cite{hore2025conformal} is similar. It differs by using a full kernel centered at $t$, which is required to be a valid density. A point, $\tilde{t}$, is then drawn randomly from that distribution. The weights are then computed in the same way as \cite{guan2023localized}, but instead of being centered at $t$, they are centered at $\tilde{t}$. With this approach, the weighted quantiles can be computed in the same way as in~\cref{sec: IPB}. They also discuss the role of the bandwidth, and how it affects the conditional coverage, in Section 4.2.2. 

Unless the weighted approach manages to put all of the mass of the treatment assignment on $t$, it cannot alone provide finite sample conditional coverage because the treatment assignment distribution is biased. With a shrinking variance (bandwidth), this bias can disappear asymptotically, but weighting on its own will often not result in good finite coverage at every treatment value. That is why combining the approach with a decent conditional quantile model is so important when the sample size is not large. 

Further work could be done to investigate how the weights introduced in \cite{huling2024independence} perform with IPB. Their weights, distance covariance optimal weights (DCOWs), are the minimizer of a dependence measure which asymptotically induce independence between the treatment and the confounders. In their supplementary material section J, they demonstrated via simulation that the penalized version of DCOWs are close to the true stabilized weights, $f_T(t)/f_{T\mid \vect{X}}(t \mid \vect{x})$, for large samples. Hence, the penalized version of DCOWs is a potential option if one wanted to use the marginal treatment distribution for $Q_T$.

\section{Analyses} \label{sec:empircal_performance}

\subsection{Simulation Studies} 
\begin{figure}[H]
    \centering
\includegraphics[scale = .25]{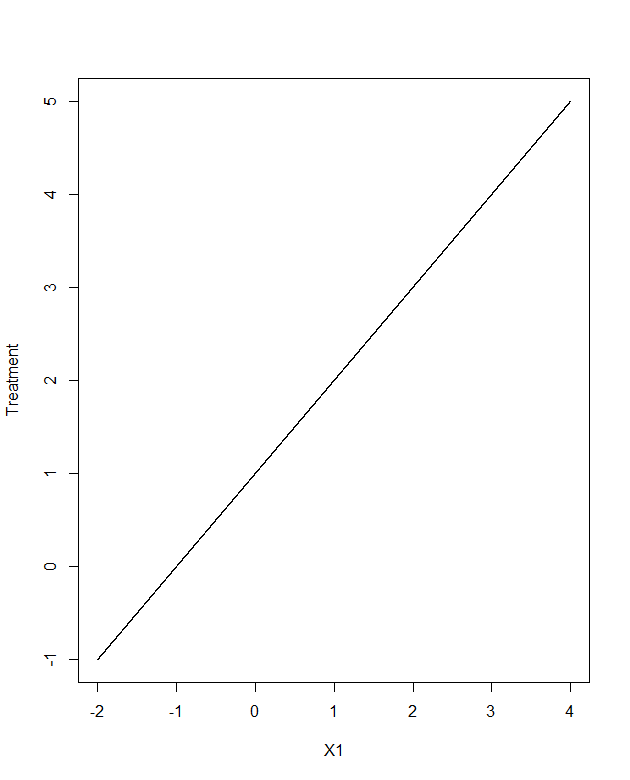}
\includegraphics[scale = .25]{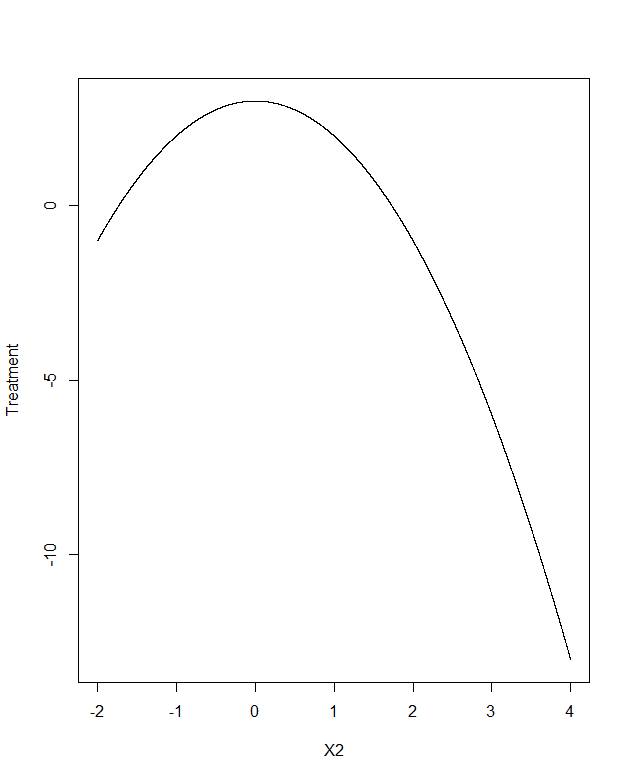}
\includegraphics[scale = .25]{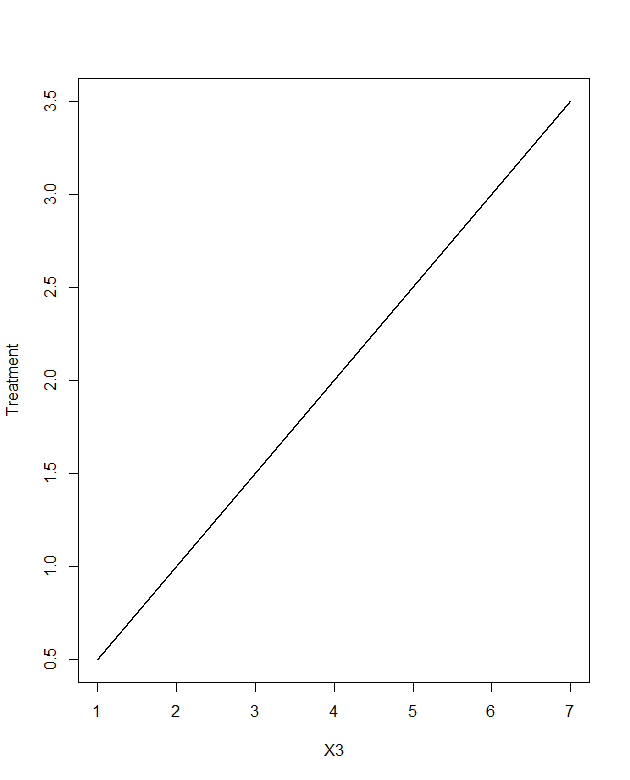}

\caption{Treatment covariate relationships for simulation scenarios 1 and 2.}\label{fig:covariate_treatment_replationship}
\end{figure}

All numerical analyses were done using the R language \citep{rlanguage}. We introduce and evaluate IPB via two different simulation scenarios. The data generation for both scenarios can be found in~\cref{tab:cov_gen}. The Normal distribution is parameterized via the mean and the variance. A visualization of the treatment covariate relationships can be found in~\cref{fig:covariate_treatment_replationship}.

\begin{table}[H]
\begin{center}
    \caption{Simulations: Data Generation}
    \label{tab:cov_gen}
    \begin{tabular}{|l|c|c|}
        \hline
        Scenario & Variable & Generating Distribution \\
        \hline 
        Both Scenarios &$X_1$ & $\mathcal{N}(1, 1)$\\
        Both Scenarios&$X_2$ & $\mathcal{N}(1, 1)$\\
        Both Scenarios&$X_3$ & $\mathcal{N}(1, 4)$\\
        Both Scenarios&$T \mid \vect{X}$ & $\mathcal{N}(X_1 - X_2^2 + 0.5X_3, 20)$\\
        Scenario 1 & $Y(t) \mid \vect{X}$ & $\mathcal{N}(X_1 + X_2 + t + X_1^2 + X_2^2 + t^2 + X_1t + X_2t + X_1X_2, 9)$\\
        Scenario 2 & $Y(t) \mid \vect{X}$ & $\mathcal{N}(X_1 + 2X_2 + t + 5X_1^2, 9)$\\
         \hline
    \end{tabular}
\end{center}
\end{table}

For both simulation scenarios, there were 5 setups that were considered. The target coverage rate was set to be 90\%. 

\begin{enumerate}
    \item Oracle outcome and oracle weight models
    \item BART outcome model and oracle weight model
    \item Oracle outcome model and estimated weights
    \item BART outcome model and estimated weights
    \item BART with no adjustment
\end{enumerate}

The oracle outcome model used the true conditional distribution of $Y(t)\mid \vect{X}$ to form the unconformalized quantiles. The Bayesian Additive Regression Trees (BART) outcome model used the 5th and 95th percentiles of the posterior predictive distribution, which was constructed from 2,000 MCMC samples computed using the \textit{dbarts} R package \citep{dbarts}. Letting $Y_i$ denote the observed response, the BART model for regression assumes the form,
\[
Y_i = f(\vect{X}_i, T_i) + \epsilon_i, \quad \epsilon_i \overset{i.i.d.}{\sim}N(0, \sigma^2),
\]
where $f$ is represented by the sum of $m$ regression trees,
\[
f(\vect{x}, t) = g(\vect{x}, t; T_1, M_1) + \cdots + g(\vect{x}, t; T_m, M_m),
\]
where $T$ is the tree structure and $M$ is the terminal node. Typically inference with BART is done through MCMC by placing priors on the parameters in the model \citep{actbart, hill2020bayesian}. More details can be found in Appendix E of the supplementary materials.

The oracle treatment weight model was the model whose conditional distribution, $T\mid \vect{X}$, was correctly specified and estimated via ordinary least squares regression. The conditional mean was estimated via ordinary least squares regression, and quantiles were formed assuming Normality. The estimated weight model used the \textit{flexmix} R package to estimate the conditional density of $T \mid \vect{X}$ with a Gaussian mixture distribution with up to 2 mixtures, with the number of mixtures selected via BIC \citep{flexmix}. The conditional mean was assumed linear in the covariates for each mixture component. The BART model with no adjustment used the estimated conditional quantiles without a conformal adjustment. 

For the treatment assignment distribution, $Q_T$, the mean was chosen to be $1$ because it is the mean of the observed treatment values when averaged over the distribution of the confounders and covariates. For the variance, $1$ was chosen because it concentrates the mass near $1$ without causing positivity problems for the treatments of interest (from 0.16 to 1.84). Using the approach discussed in~\cref{sec:treatment assignment}, $h = 1$ can be thought of as the maximum value of $h_k$ that we saw over all the simulations and treatment values with a small amount added to ensure positivity. In practice, $h$ can be picked to be the maximum over the treatment values of interest after checking for positivity. 

The simulation size for both scenarios was $N = 1,000$. In both simulation scenarios, the observed sample size was $n = 1,000$, which was split equally between a training and a calibration set. Within each simulation, one out of sample covariate was generated, $\vect{x}_{\text{oos}}$, according to the same distribution as the observed data. Ten treatments of interest were then used. These were chosen by computing 10 equally spaced percentiles from 20\% to 80\%. The corresponding treatment of interest was then found assuming the percentiles corresponded to $\mathcal{N}(1, 1)$. For example, the first treatment of interest percentile was the 20th percentile, which corresponds to an actual treatment of $0.16$. The final treatment corresponding to the 80th percentile was $1.84$. The weights were computed assuming the assignment (shifted) distribution, $Q_T$, was $\mathcal{N}(1, 1)$. Ten potential outcomes were then generated using the distributions from~\cref{tab:cov_gen} with $\vect{x}_{\text{oos}}$ and the ten treatments of interest. Average coverage and length can be found in~\cref{tab:sim1_results} and~\cref{tab:sim2_results}. Plots of conditional coverage for both simulation scenarios can be found in~\cref{fig:cond_cov}.

\begin{table}[H]
\begin{center}

    \caption{Simulation Result: Simulation 1}
    \label{tab:sim1_results}

    \begin{tabular}{|c|c|c|c|}
        \hline
         Situation &  Coverage & Length \\
         \hline
         Oracle model \& oracle weights & 0.900 (0.003) & 9.891 (0.026) \\
        BART outcome model \& oracle weights  & 0.903 (0.004) & 12.975 (0.124)  \\
         Oracle outcome model \& estimated weights &  0.901 (0.003) & 9.903 (0.024) \\
         BART outcome model \& estimated weights  & 0.902 (0.004) & 12.848 (0.121)  \\
        BART outcome model \& no adjustment & 0.722 (0.006) & 8.601 (0.117) \\
         \hline
    \end{tabular}
\end{center}
\end{table}

\begin{table}[H]
\begin{center}

    \caption{Simulation Result: Simulation 2}
    \label{tab:sim2_results}

    \begin{tabular}{|c|c|c|c|}
        \hline
         Situation &  Coverage & Length \\
         \hline
         Oracle model \& oracle weights & 0.900 (0.003) & 9.891 (0.026) \\
        BART outcome model \& oracle weights  & 0.897 (0.003) & 10.919 (0.064)  \\
         Oracle outcome model \& estimated weights &  0.901 (0.003) & 9.903 (0.024) \\
         BART outcome model \& estimated weights  & 0.898 (0.003) & 10.958 (0.063)  \\
        BART outcome model \& no adjustment & 0.494 (0.006) & 4.519 (0.058) \\
         \hline
    \end{tabular}
\end{center}
\end{table}

\begin{figure}[H]
    \centering
\includegraphics[scale = .35]{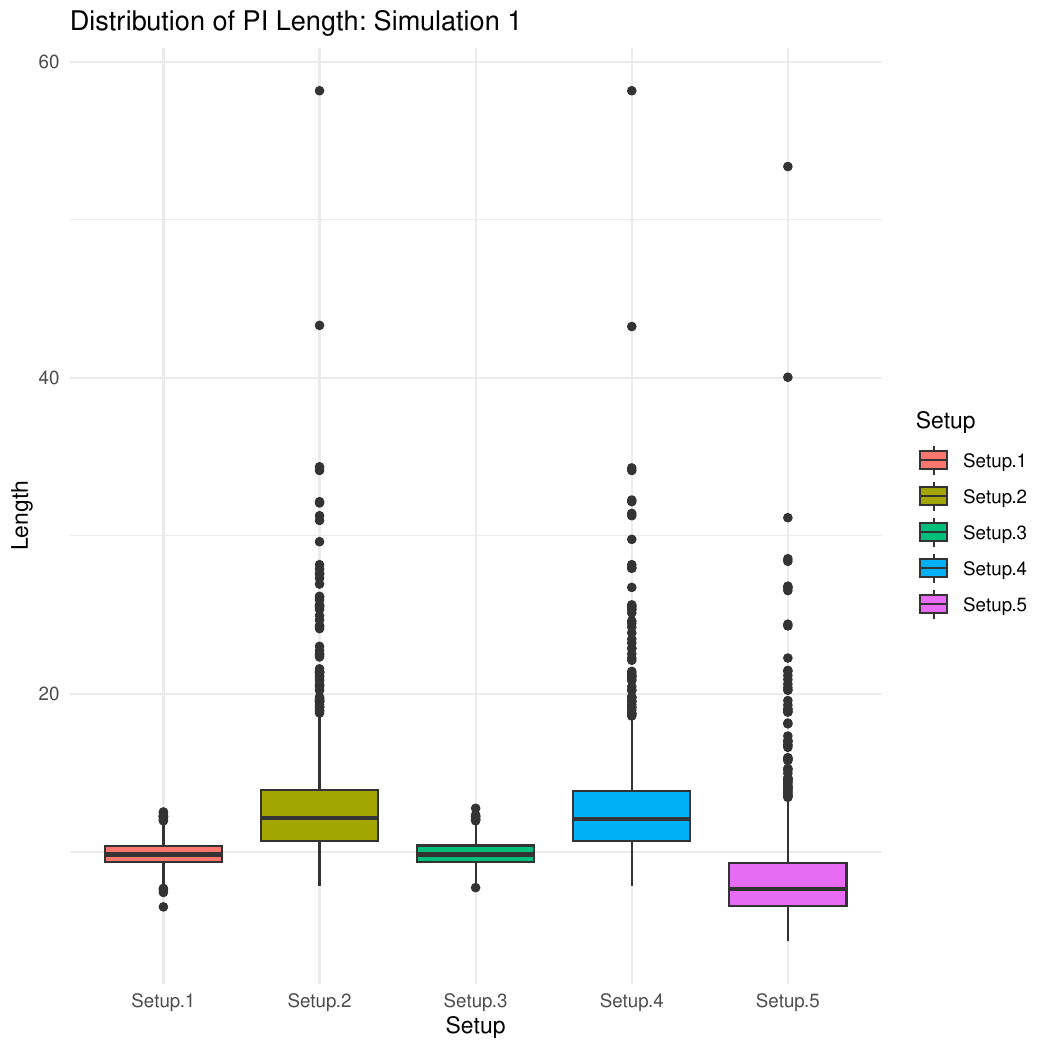}
\includegraphics[scale = .35]{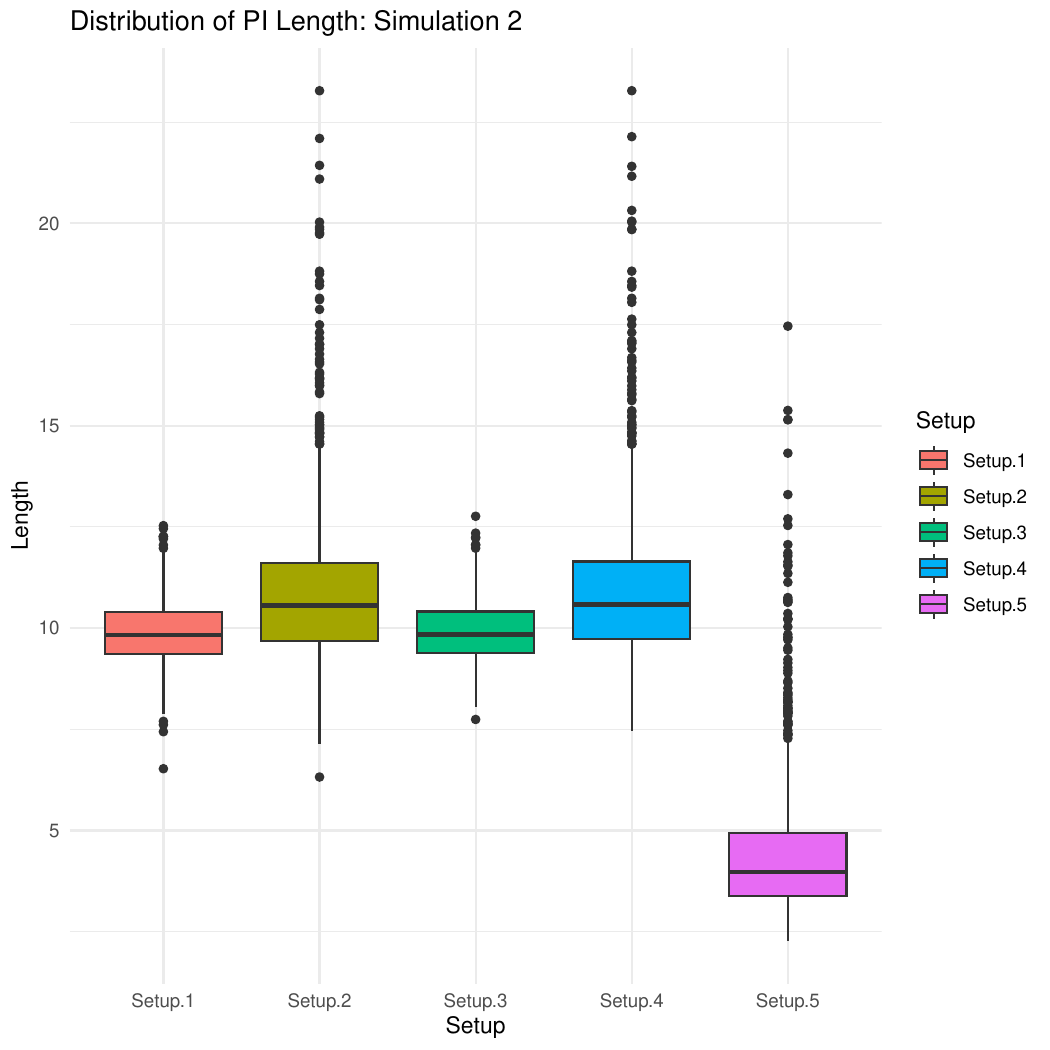}
\caption{The image on the left (right) is the distribution of the prediction interval lengths from the first (second) simulation. }\label{fig:sim_length_comp}
\end{figure}

\begin{figure}[H]
    \centering
\includegraphics[scale = .35]{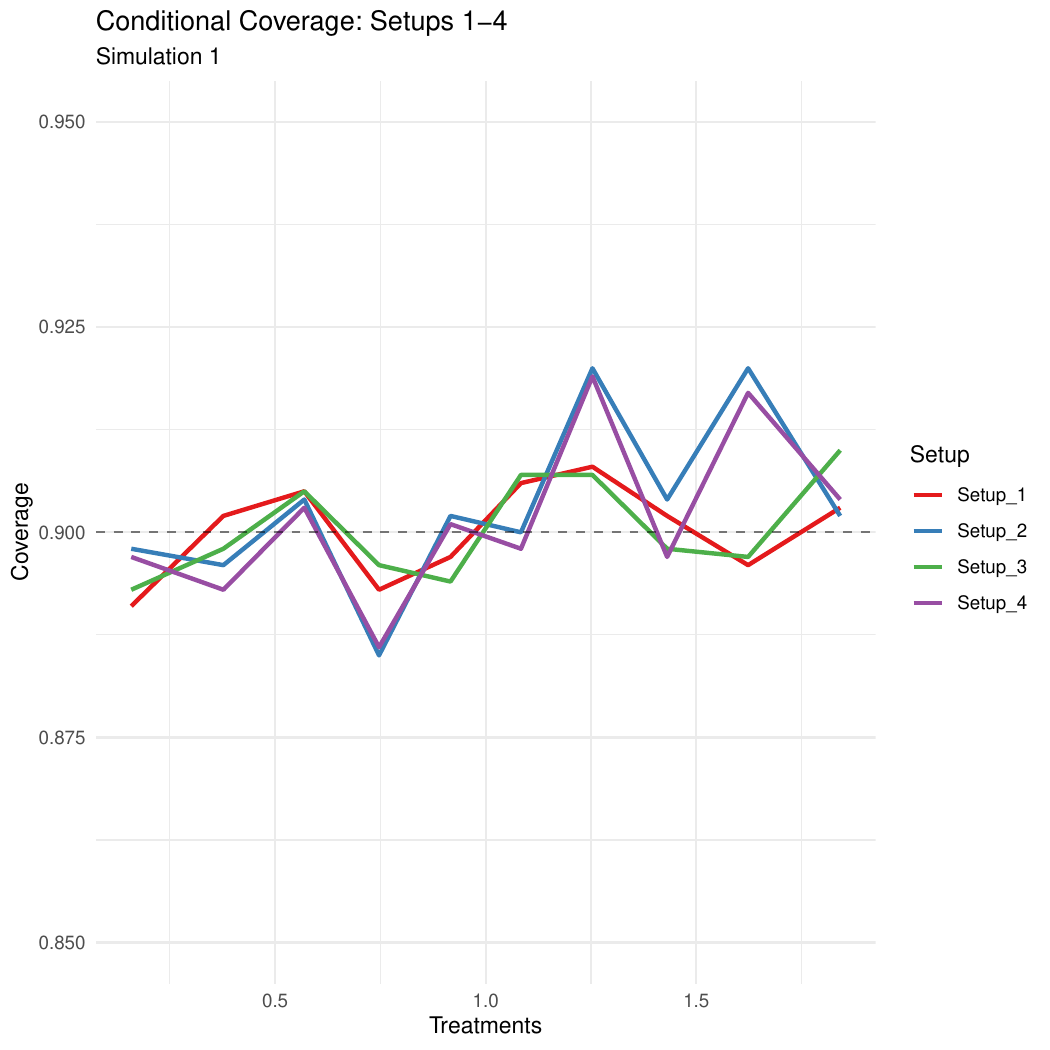}
\includegraphics[scale = .35]{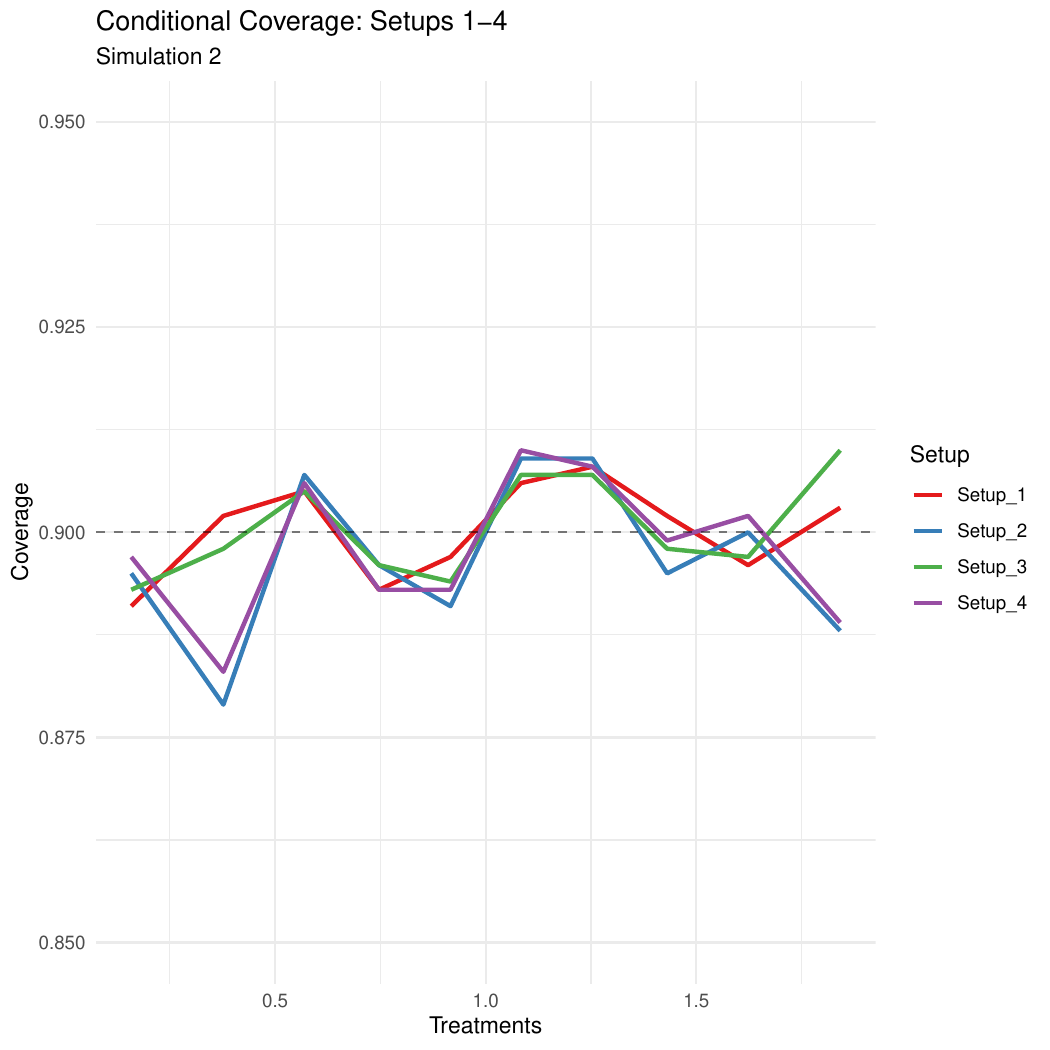}
\caption{The image on the left (right) is the empirical conditional coverage from the first (second) simulation.  Setup 5 is not included in these comparisons because its conditional coverage is off the plot. }\label{fig:cond_cov}
\end{figure}

It is clear from both scenarios that an unadjusted conditional quantile model is not a reasonable model to use (setup 5). From the other four setups, we can see that the length of the prediction interval depends on the conditional quantile model used. While the method is asymptotically doubly robust, sharp (small) prediction intervals (and hence bands) with nominal finite sample coverage require a well specified outcome regression model that carefully takes the structure of the response and treatment relationship given the confounders into account. This can be seen by looking at the distribution of the prediction interval lengths in~\cref{fig:sim_length_comp}, where the prediction bands formed using the oracle outcome regression model are both smaller and less variable than those produced using the BART outcome regression model. 

We can also see from the conditional coverage plots in~\cref{fig:cond_cov} that, although conformal prediction only guarantees marginal coverage, conditional coverage (and, hence, prediction bands with valid coverage over all reasonable treatment values) is a reasonable assumption when combining weighted conformal prediction with quantile regression.

\subsubsection{Treatment Truncation} \label{sec:treat_trunc}

The motivation for this numerical experiment was given in~\cref{sec:methods}. Two scenarios were looked at, one with a homoskedastic error and one with a heteroskedastic error. One covariate, $X$, was generated from a standard Normal distribution. The treatment given the covariate, $T\mid X$, was generated from a truncated Normal distribution with a mean of $X^2 + 1$. In both scenarios, it was truncated to only take values between 0.5 and 5. In the homoskedastic case, the variance was one. In the heteroskedastic case, the variance was $X^2$. The conditional response, $Y\mid X, T$ was generated from a Normal distribution with a mean of $3X + T + X\times T$. The sample size for each simulation was $n = 10,000$. This was split equally between a training set and a calibration set. Quantile regression and the quantile score function were used. The conditional quantiles were correctly specified and estimated from the data. The simulation number was $N = 1,000$. In each simulation, ten additional points were generated. The covariate, $X$, and the conditional response, $Y\mid T, X$, were generated in the same way. The treatment was generated from a truncated Normal distribution with a mean of 2 and a variance of 0.8 for both scenarios. It was truncated to take values between 1 and 5. The oracle weights were used and the denominator of the weights was given a slight offset of 0.001. The offset was needed because positivity is not guaranteed in each simulation, so 0.001 is added to the GPS to avoid extremely large weights, and infinite prediction intervals. The treatment assignment distribution was correctly specified in both scenarios. In practice, positivity should be checked for to ensure valid causal claims, so no offset is required. 

\begin{table}[t]
\begin{center}
     \caption{Simulation Result: Truncated Treatment}
    \label{tab:sim_trunc_trt}

    \begin{tabular}{|c|c|c|}
        \hline
         Scenario & Coverage & Length \\
         \hline
         Homoskedastic & 0.950 (0.002) & 11.788 (0.186)\\
         Heteroskedastic & 0.949 (0.003) & 11.796 (0.187)\\
         \hline
    \end{tabular}
\end{center}
\end{table}

Here, as with the previous simulation studies, the method works especially well when both the outcome model and the weight model are correctly specified, see Table~\ref{tab:sim_trunc_trt}.

\subsection{Real Data Analysis}

 The data comes from the 1987 National Medical Expenditure Survey (NMES). It was originally extracted and analyzed by \cite{johnson_elizabeth_2003} and analyzed again by \citet{imaidyke2004, galagate2016causal}. The data can be found in the R package \textit{causalrdf} \citep{causaldrf}. Our analysis includes the following subject-level covariates: age at the times of the survey (19–94), age when the individual started smoking, gender (male, female), race (white, black, other), marriage status (married, widowed, divorced, separated, never married), education level (college graduate, some college, high school graduate, other), census region (Northeast, Midwest, South, West), poverty status (poor, near poor, low income, middle income, high income), and seat belt usage (rarely, sometimes, always/almost always). The response of interest was total medical expenditure (measure in 1987 US dollars). The treatment of interest was \textit{packyear}. It combines self-reported information about frequency and duration of smoking \citep{johnson_elizabeth_2003}.
\[
\textit{packyear} = \frac{\text{number of cigarettes per day}}{20} \times \text{number of years smoked}
\]
Following \cite{imaidyke2004, galagate2016causal}, we use log(\textit{packyear}) as the treatment variable. This allowed us to comfortably use a Normal distribution to estimate the GPS. 

Covariate balance or overlap was checked with the method proposed in \cite{flores2012} and implemented in the R package \textit{causaldrf} \citep{causaldrf}. To achieve positivity, 1944 of the 9708 observations were removed. This left us with 7764 observations, which were split evenly into a calibration set and a training set. Selected plots of the covariate distributions can be found in Appendix A of the supplementary materials.

We follow \citet{galagate2016causal} in looking at the population average dose-response curve along with a 95\% confidence interval, though we use the method introduced by \citet{HiranoandImbens2004}. The population dose-response curve along with a 95\% bootstrapped confidence interval can be found in~\cref{fig:adr_ci}. A description of how the population dose-response curve was fit is described in Appendix D of the supplementary materials. The confidence interval is shaded in gray, while the curve is given by the dashed line. There are 250 randomly selected individuals whose log \textit{packyear
} values are plotted vs their total medical expenditure as points. From this we can see that the individual variability is not well captured by a population average dose-response curve and corresponding confidence interval, as expected. 

Next, we randomly selected 5 individuals who were older than 22 and younger than 45 in 1987 to ask the causal question, what is your 90th percentile total medical expenditure if you stop smoking now versus your 90th percentile total medical expenditure if you continue smoking at 80\% of your current rate? We computed 5 prediction intervals for each individual in 3 year increments, and combined them to form prediction bands in both scenarios. For example, if an individual had a packyear level of $10$, had been smoking for 10 years, and was 30 at the time of the study, the first prediction interval computed would assume they are now 33 with a packyear level of $10 \text{ (the current packyear)} + 1 \text{ (the average rate their packyear increases)} \times 0.80 \text{ (80\% of the current rate)} \times 3 \text{ (three year increments)} = 12.40$. We then use the natural log of $12.40$ as the treatment of interest to form the first prediction interval for this individual. Both the log(packyear) and age change for each prediction interval, while all other confounders remain the same. This process is repeated 5 times for each individual before the prediction intervals are combined to form prediction bands. 

Only the upper 90th percentile of total medical expenditure is investigated because the 10th percentile of the unconditional response is \$0. That is, the majority of the time the lower-bound would be $\approx$ \$0, and would not be different or of interest to the majority of individuals. We also would like to convince the smokers in this hypothetical situation that they can save money by quitting smoking, so we look at the worst-case: if they keep smoking, which corresponds to looking at the one-sided prediction bands.

Those comparisons can be found in~\cref{fig:ind1}. The covariates for these five individuals can be found in Tables \ref{tab:ind_covar} - \ref{tab:ind_covar2}. All individual prediction models were computed using BART. The data were split equally into a training set for the model, and a calibration set to compute non-conformity scores. 

The prediction bands for continued smoking were formed assuming that if the individual continued to smoke, their packyears would increase at 80\% of their current smoking rate. This was done to help avoid extreme levels of smoking, which would have caused issues with extrapolation. If that rate led to them having the highest packyear in the dataset, we truncated their packyear to be equal to the highest observed in the dataset. This was only the case for the second individual's last prediction interval.

To form the treatment assignment distribution, we first broke the levels of log packyear into deciles. The assignment distribution was then a Normal distribution with a mean equal to the mid-point of the treatment of interest's decile and a standard deviation equal to the standard deviation used to estimate the GPS. The decile approach was used because the packyear variable was self-reported by individuals, so we hypothesize that treatments near each other (in the same decile) are very similar. Therefore, we introduce this discretization to ensure our results are robust. An approach that uses an even larger weight for individuals in the calibration set with a similar packyear to the treatment of interest in shown in Appendix C of the supplementary materials. This corroborates that discretizing the treatment into deciles is robust to slight changes in the treatment assignment distribution. A density plot of the marginal treatment density, along with a plot of the treatment assignment distribution, can be found in~\cref{fig:pkyrs_dens}.

Formulaically, the weights used on the scores for a treatment of interest $t$ are:
\[
W_i = 
\frac{\phi(T_i \mid d(t), s^2)}{\phi(T_i \mid \hat{\mu}(\vect{X}_i), s^2)}, 
\]
where $d(t)$ is the mid-point of the decile containing $t$,
$\phi(\cdot\mid a, b)$ denotes the density of a Normal distribution with a mean of $a$ and a variance of $b$, $\hat{\mu}(\vect{x})$ denotes the estimated mean from the GPS, and $s^2$ the mean squared error from the GPS model. 

For each of the five individuals we look at five levels of the potential treatment (smoking level measured by the log of the packyears). For simplicity we ignore the log scale to explain how the treatment and age vary in our analysis. We can view the individual's treatment of interest (the hypothesized packyear) as a function of their age at the baseline plus the change in their age,
\[
t(\text{age}_0 + \delta) = t_0 + r\delta,
\]
where $t_0$ is the baseline packyear, $r$ is 80\% of their current rate of smoking, and $\delta$ is the number of years since 1987 ($\delta \in \{3, 6, 9, 12, 15\}$ in our analysis). This also allows us to represent the mid-point of the decile of the treatment of interest,
\[
d(t(\text{age}_0 + \delta)),
\]
Clearly this allows us to incorporate the changes in packyear and age into both the outcome regression model and the weights for the non-conformity scores. 

\begin{table}[t]
\begin{center}

    \caption{Selected Individual's Covariates}
    \label{tab:ind_covar}

    \begin{tabular}{|c|c|c|c|c|}
        \hline
         Age (1987) &  Age when smoking began & Gender & Race & packyear (1987)\\
         \hline
         35 & 21 & Female & White & 22.5\\
         40 & 19 & Male & White & 38.5 \\
         42 & 19 & Male & White & 11\\
         36 & 15 & Female & White & 16.5\\
         39 & 17 & Female & White & 34.5\\
         \hline
    \end{tabular}
\end{center}
\end{table}

\begin{table}[t]
\begin{center}

    \caption{Selected Individual's Covariates}
    \label{tab:ind_covar2}

    \begin{tabular}{|c|c|c|c|c|}
        \hline
         Marriage & Education & Census Region & Poverty & Seat Belt\\
         \hline
         Never Married & High School & West & Middle Income & Rarely \\
         Married & Some College & Midwest & High Income & Always \\
         Never Married & High School & West & High Income & Rarely \\
         Married & Some College & Midwest & Middle Income & Always \\
         Never Married & Some College & Midwest & Middle Income & Always \\
         \hline
    \end{tabular}
\end{center}
\end{table}

\begin{figure}[H]
    \centering
\includegraphics[scale = .5]{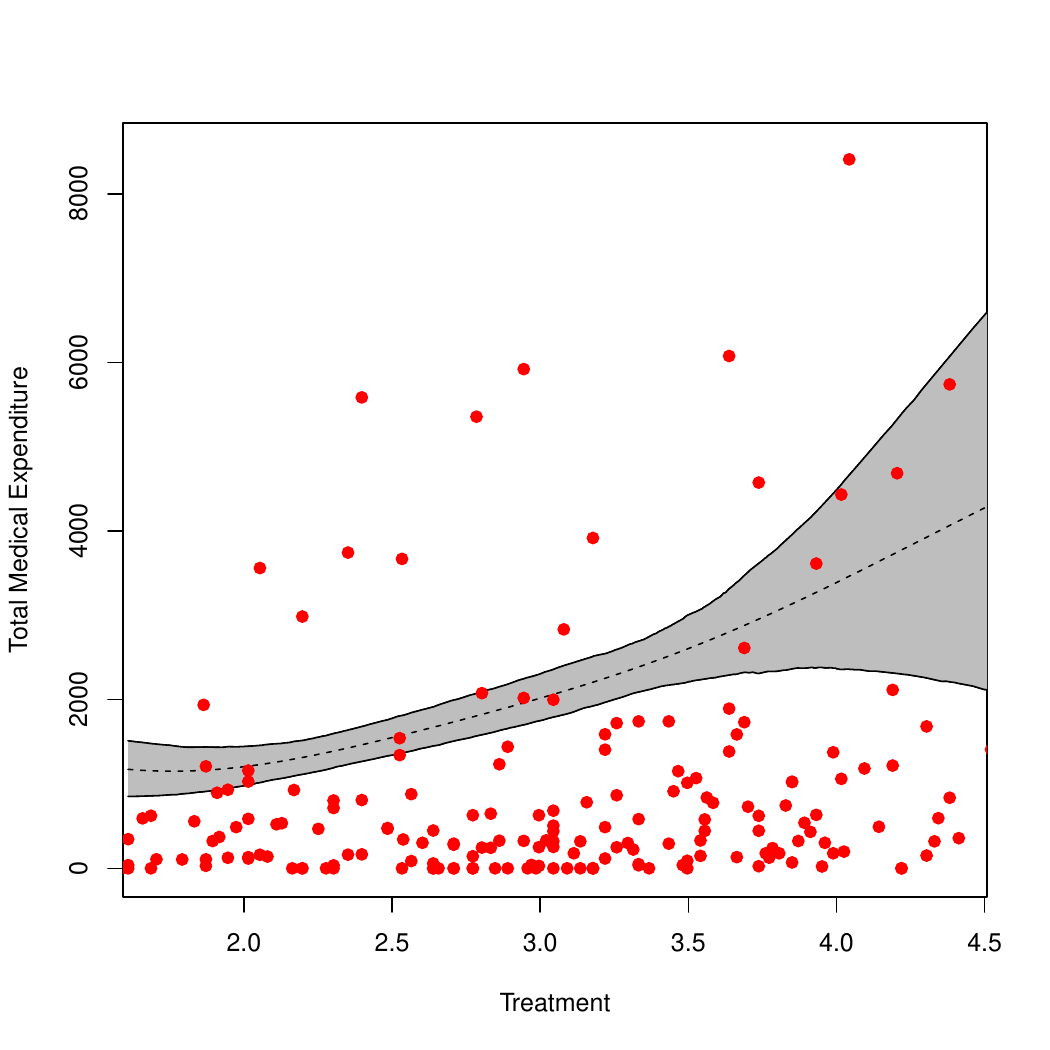}
\caption{Average dose-response curve and 95\% confidence band. Points are randomly selected participant's observed total medical expenditure.}\label{fig:adr_ci}
\end{figure}

\begin{figure}[t]
    \centering
\includegraphics[scale = .3]{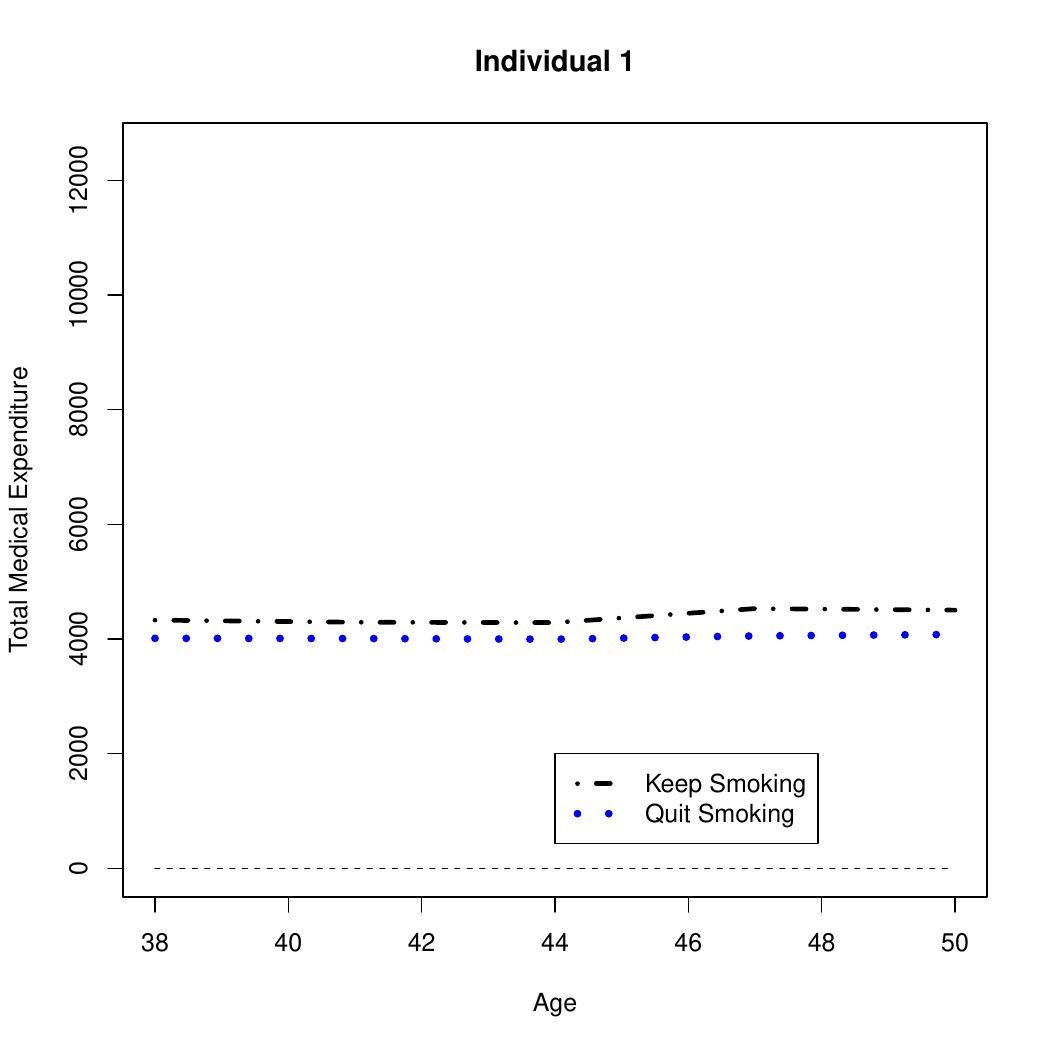}
\includegraphics[scale = .3]{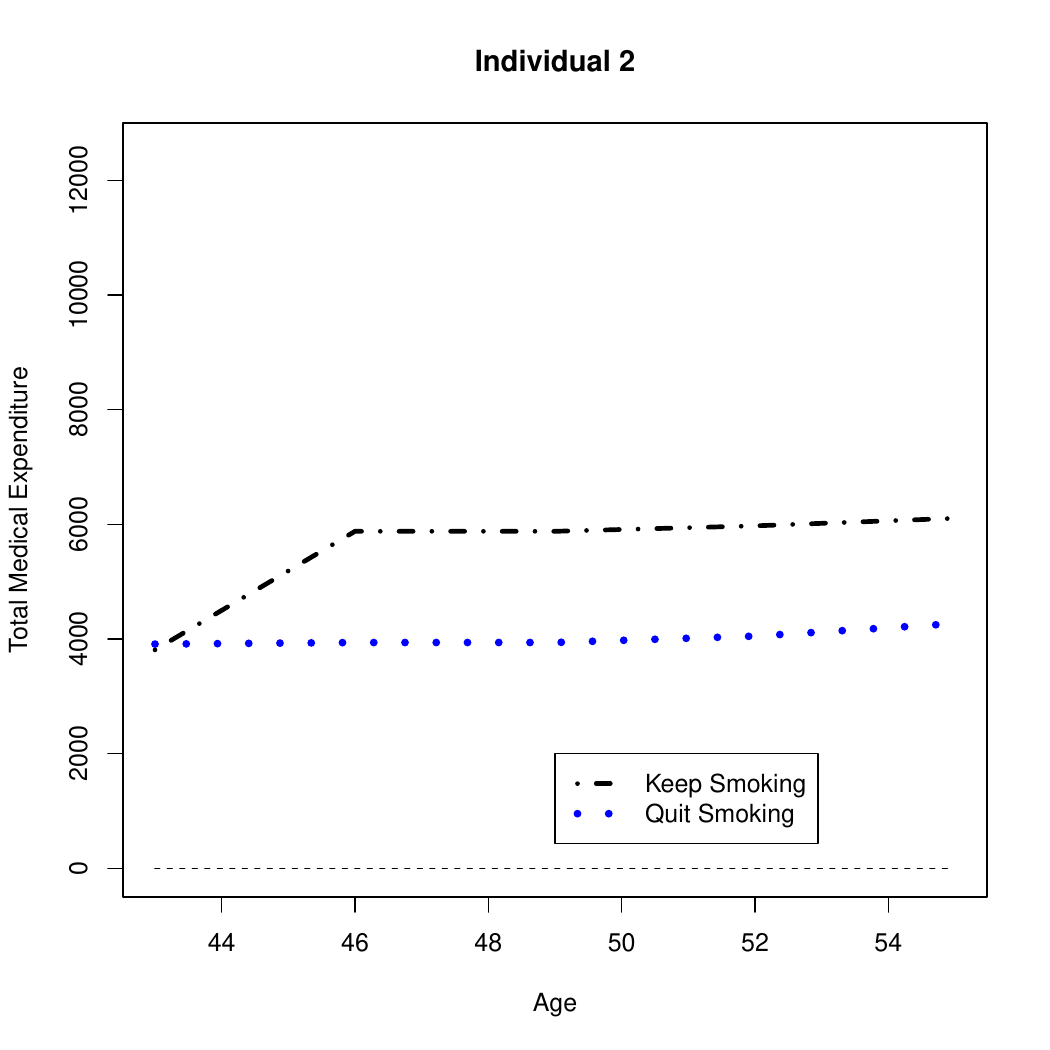}
\includegraphics[scale = .3]{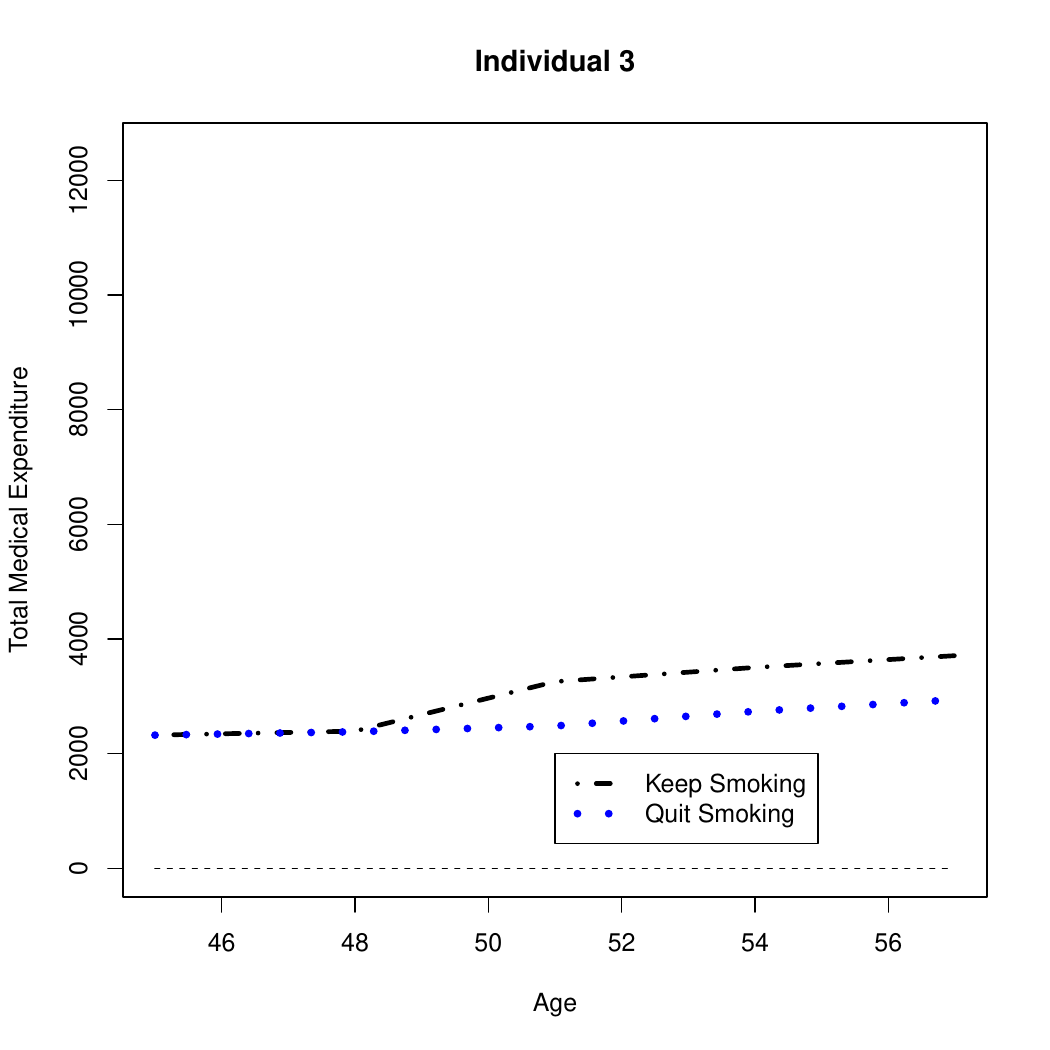}
\includegraphics[scale = .3]{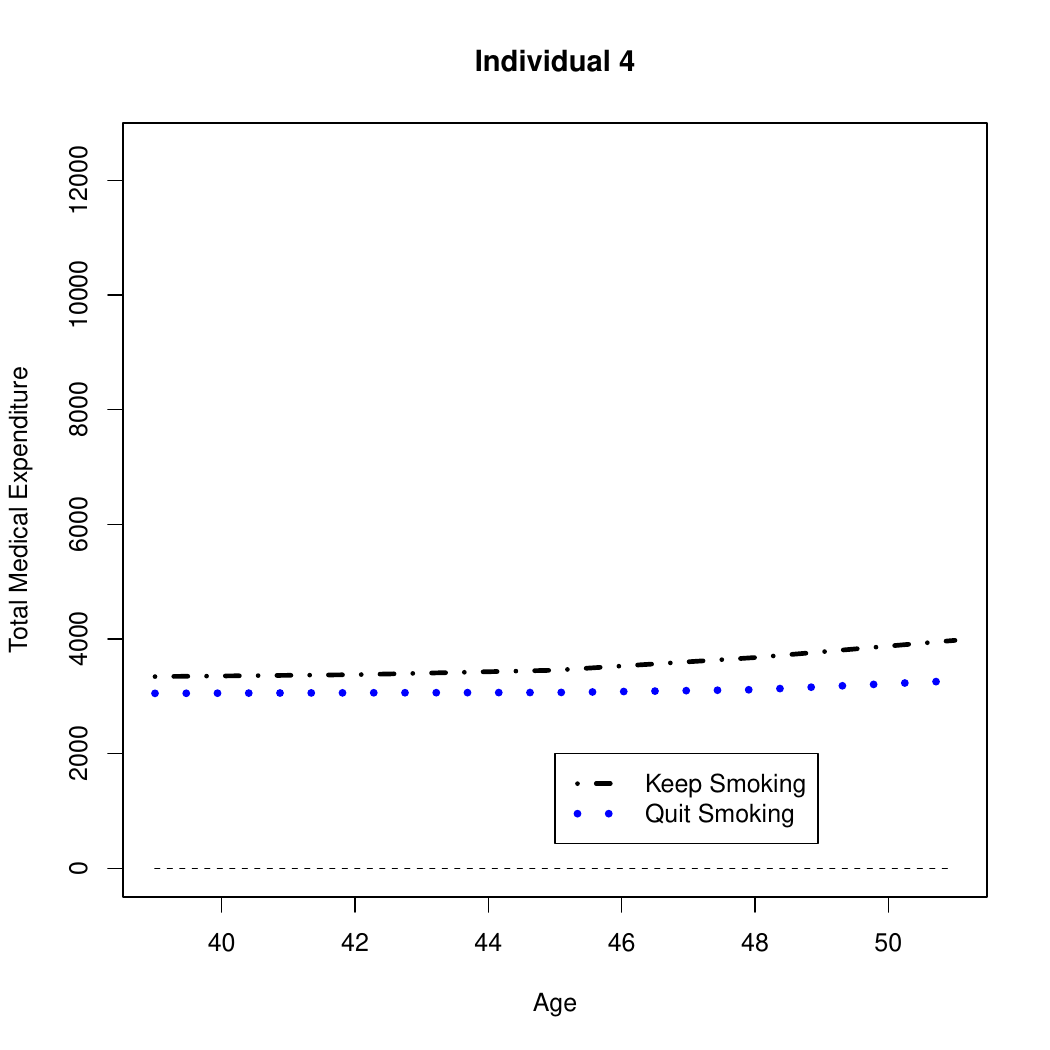}
\includegraphics[scale = .3]{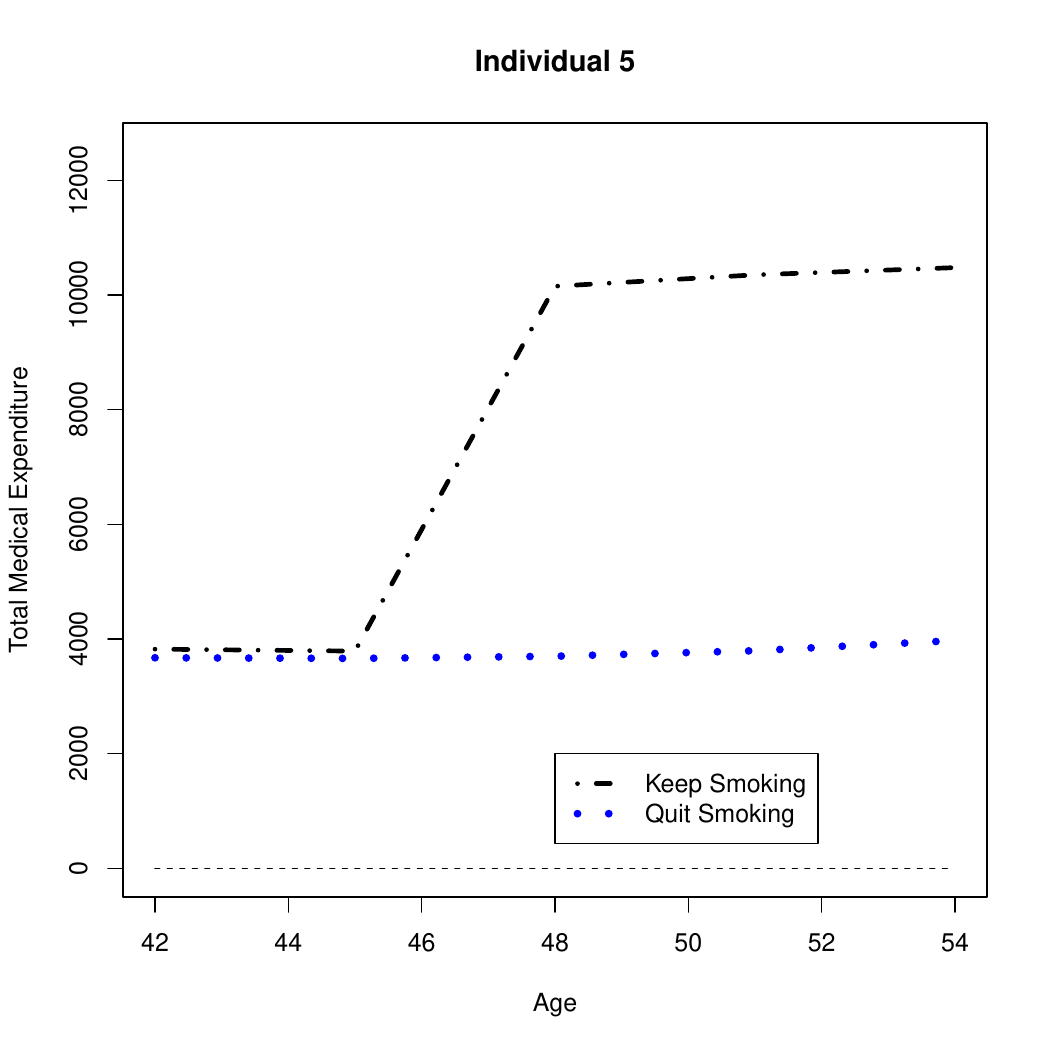}

\caption{90\% upper conformal prediction intervals for continued smoking are given by the black dashed lines. 90\% upper prediction intervals for quitting smoking are given by the blue line. }\label{fig:ind1}
\end{figure}

\begin{figure}[t]
    \centering
    \subfloat[\centering A marginal density plot of the log(packyear) variable, along with the selected individual's packyear value plotted.]{\includegraphics[width = 6.5cm]{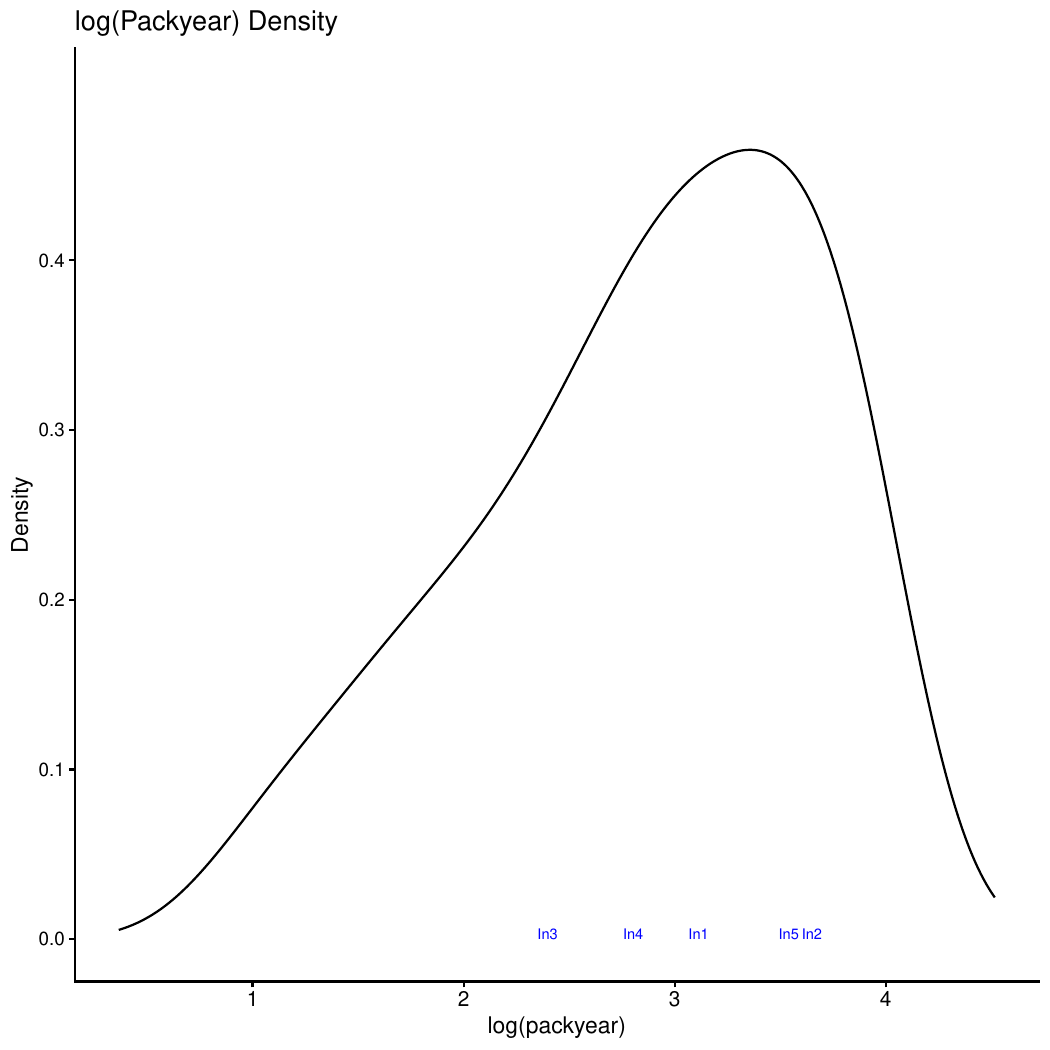}}
    \qquad 
    \subfloat[\centering The treatment assignment density with a decile midpoint equal to the starting level for the third individual.]{\includegraphics[width = 6.5cm]{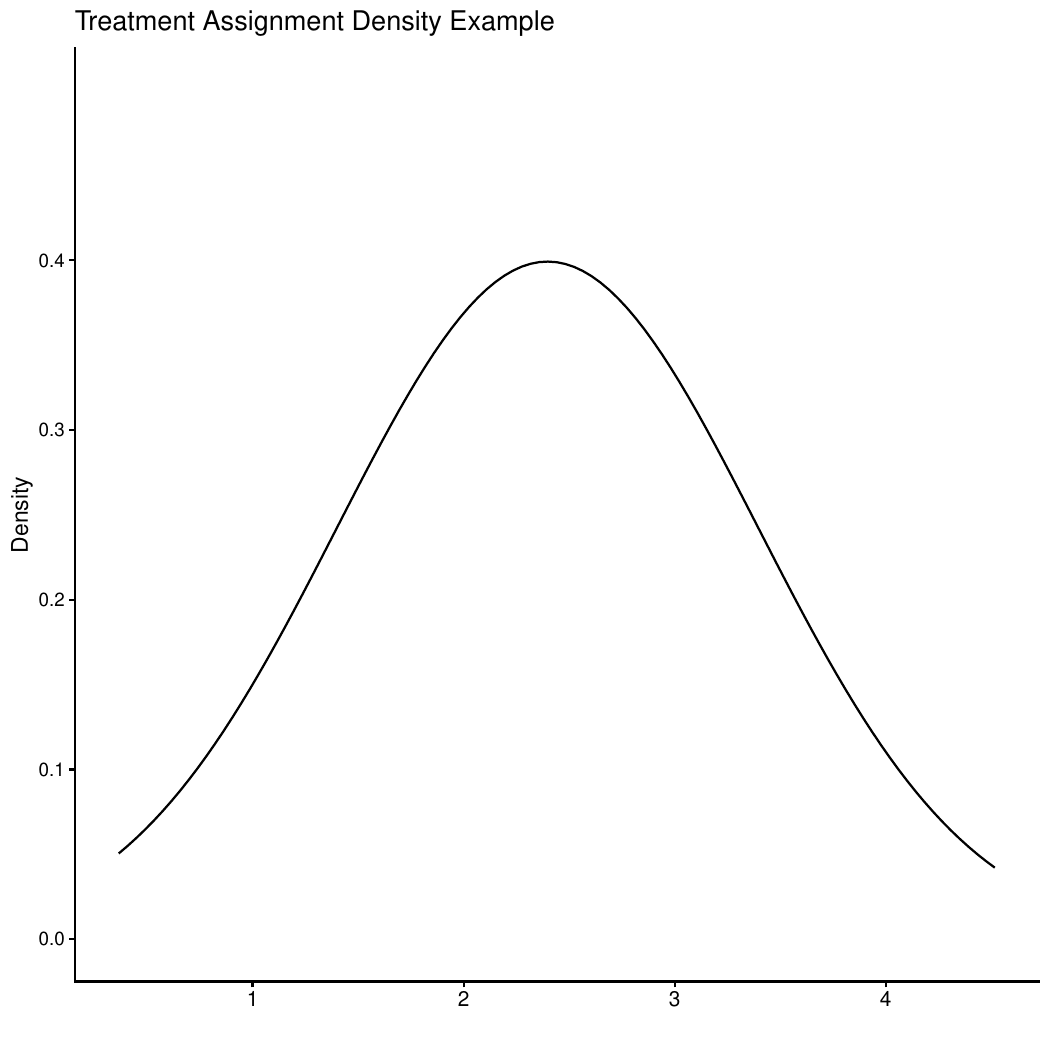}}
\caption{Treatment Distributions (on the same scales)}\label{fig:pkyrs_dens}
\end{figure}

From the plots in~\cref{fig:ind1}, we can see the general pattern is that if an individual continues to smoke, their total medical expenditure will increase. These plots can be used to show individuals who smoke how much they could save over their lifetime, and that there is still a benefit to be had by quitting smoking. For certain individuals, such as the 5th individual, continuing to smoke may greatly increase their total lifetime medical expenditure. Beyond individuals, these costs are also relevant to insurance companies and governments, which fund programs to help individuals quit smoking or offer lower premiums to those who quit smoking \citep{izumi_smoking_cost_japan_2001, curry1998_smoking_cessation_insurance, fishman2003health}. 

Furthermore, these plots illustrate that the total lifetime medical expenditure will generally increase faster for those who continue to smoke than for those who quit. These differences are likely to be larger for older individuals, similar to those seen for the second and fifth individual. The third individual is a similar age to those two, but has a much lower cumulative smoking load. We would expect to see similarly large differences for the first, third, and fourth individual if we looked further into the future, but we avoid doing so to avoid extreme extrapolation.

\section{Discussion}
From the numerical results, we can see that the individual causal quantile bands have valid prediction coverage when the treatment is randomly assigned according to an assignment distribution, and not according to the observed confounder conditional distribution. They are also easy to visualize, allowing practitioners to easily see the difference between possible treatments, and optimal treatment windows.  

IPB adds uncertainty quantification with minimal assumptions to individualized dose-response curves. This method is unique compared to the existing methods because it starts with a quantile estimate, instead of a conditional mean estimate, a distinction that results in a doubly robust coverage property. Furthermore, the framework is flexible enough to allow practitioners to evaluate targeted treatment assignment policies. To support this, we developed a data-driven approach to finding parameters of the treatment assignment distribution to ensure reasonable treatment conditional coverage. Further work can be done exploring conditional treatment assignment models, instead of the unconditional treatment assignment models that we focused on.

\newpage
\bibliographystyle{apalike} 
\bibliography{ref}
\newpage 

\appendix 

\section*{Appendix of IPB}
{The appendix of IPB contains five sections. The first contains distributional plots of some of the variables used in the real data analysis. The second section looks at comparison between three treatment assignment distribution approaches and how they compare to the oracle assignment distribution. The third conducts the real data analysis with a different treatment assignment distribution that more heavily weights individuals in the calibration set who have a treatment in the same decile as the treatment of interest. The fourth reviews the generalized propensity score. The final section is a discussion of BART and its priors. }

\section{Real Data Plots}\label{sec:real_dat_plots}

\begin{figure}[H]
    \centering
\includegraphics[scale = .3]{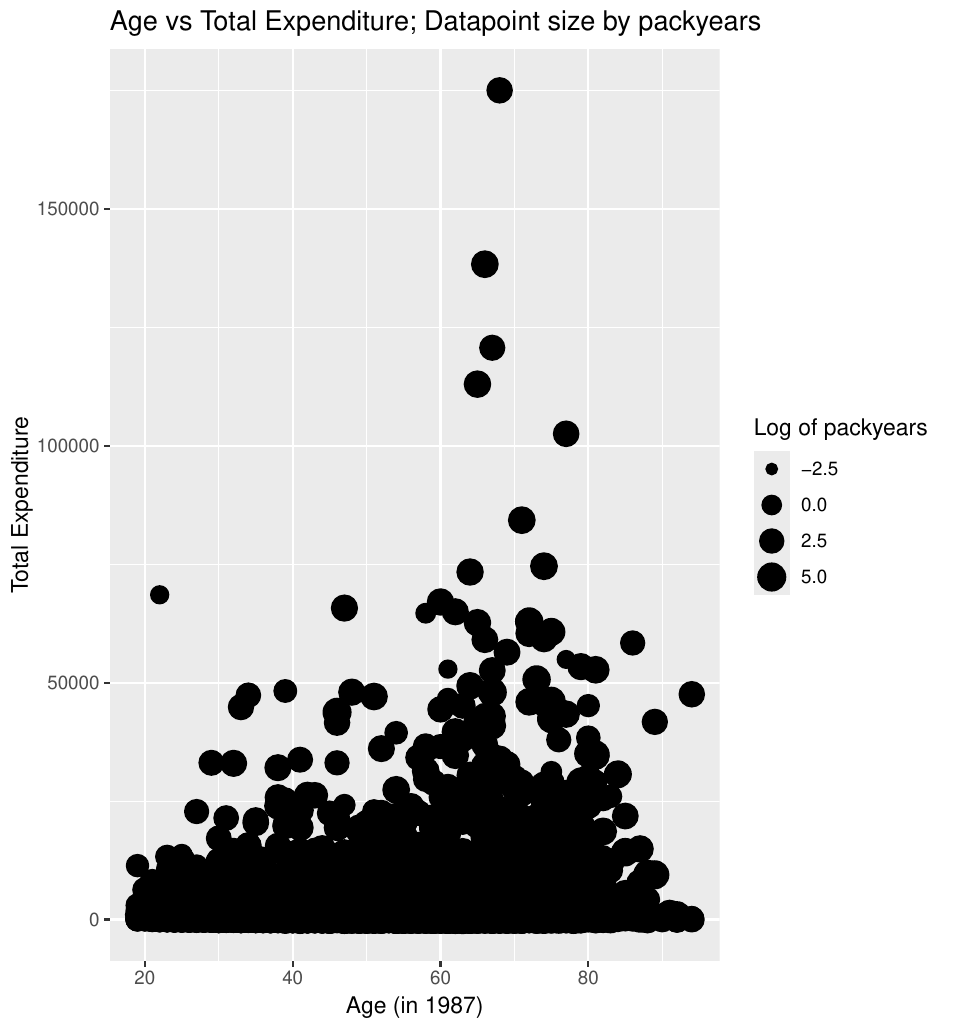}
\includegraphics[scale = .3]{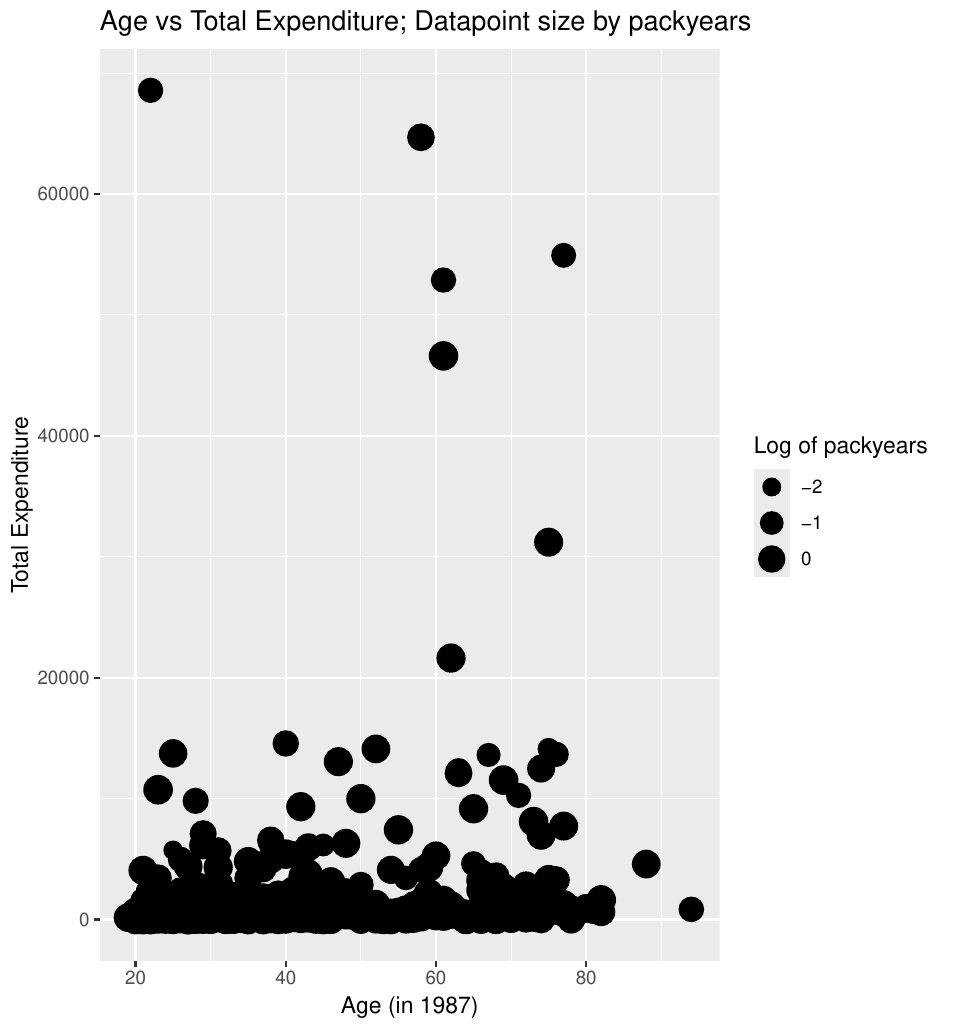}
\includegraphics[scale = .3]{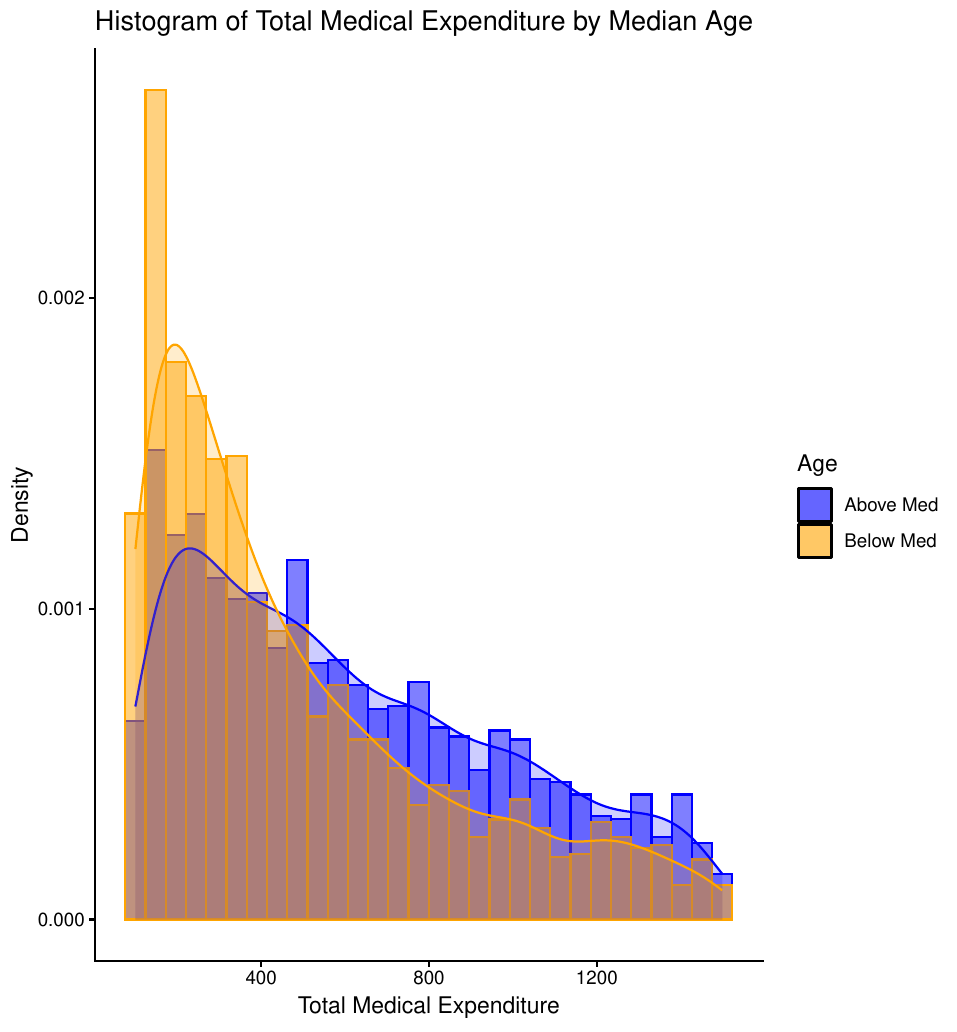}
\caption{The first plot is the full data, the second is a subsample. These plots don't appear to provide very much information because there is a point mass at \$0 for expenditure. The last plot is the distribution of total medical expenditure for those older than 48 and younger than 48. Total medical expenditure was truncated to be between the first and third quartile (\$100 and \$1500) because of the large point mass at zero, as well as extreme outliers.}\label{fig:expense_age_trt}
\end{figure}

\begin{figure}[H]
    \centering
\includegraphics[scale = .3]{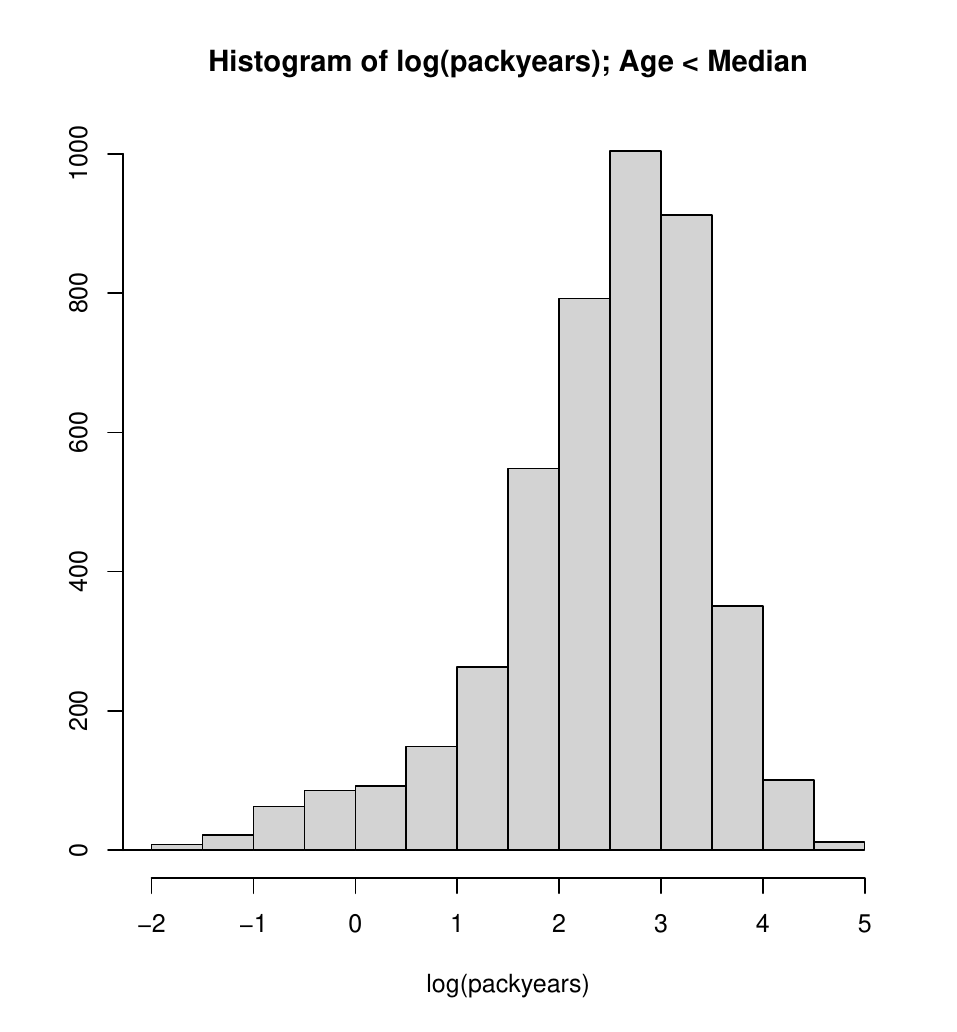}
\includegraphics[scale = .3]{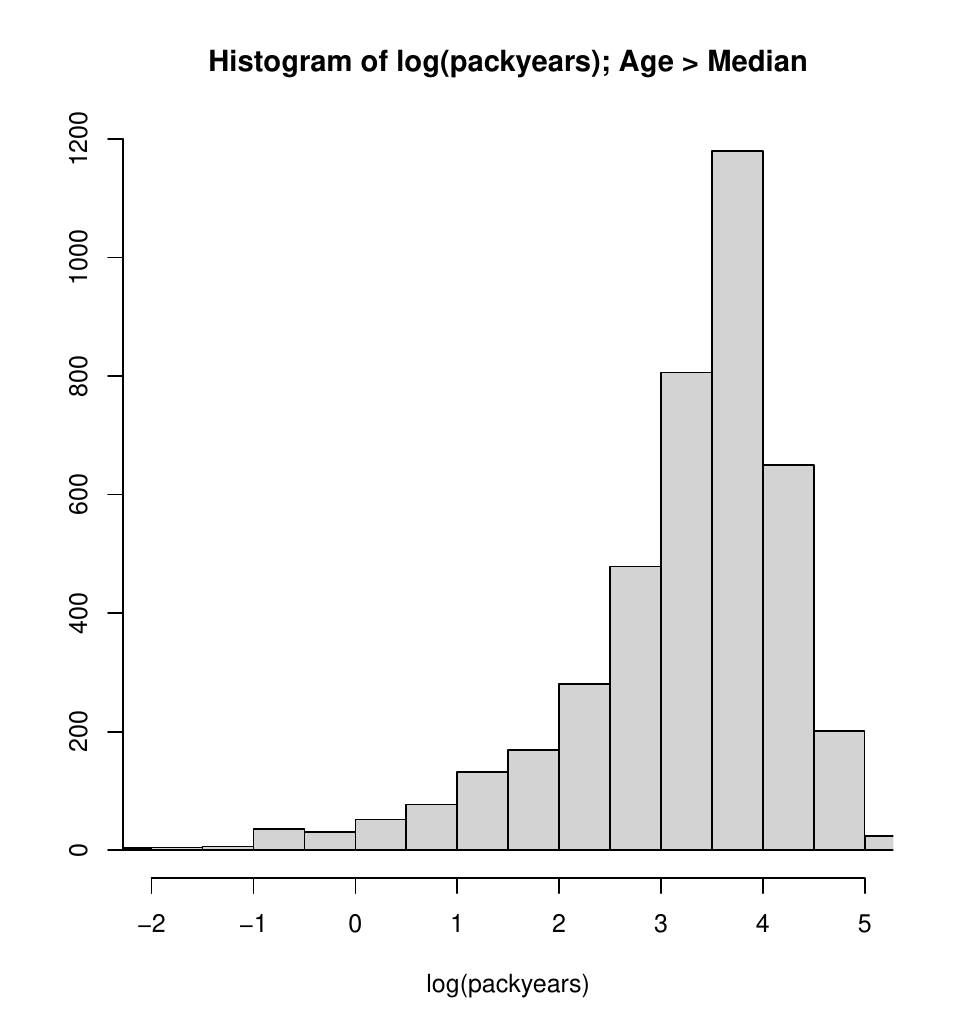}
\includegraphics[scale = .3]{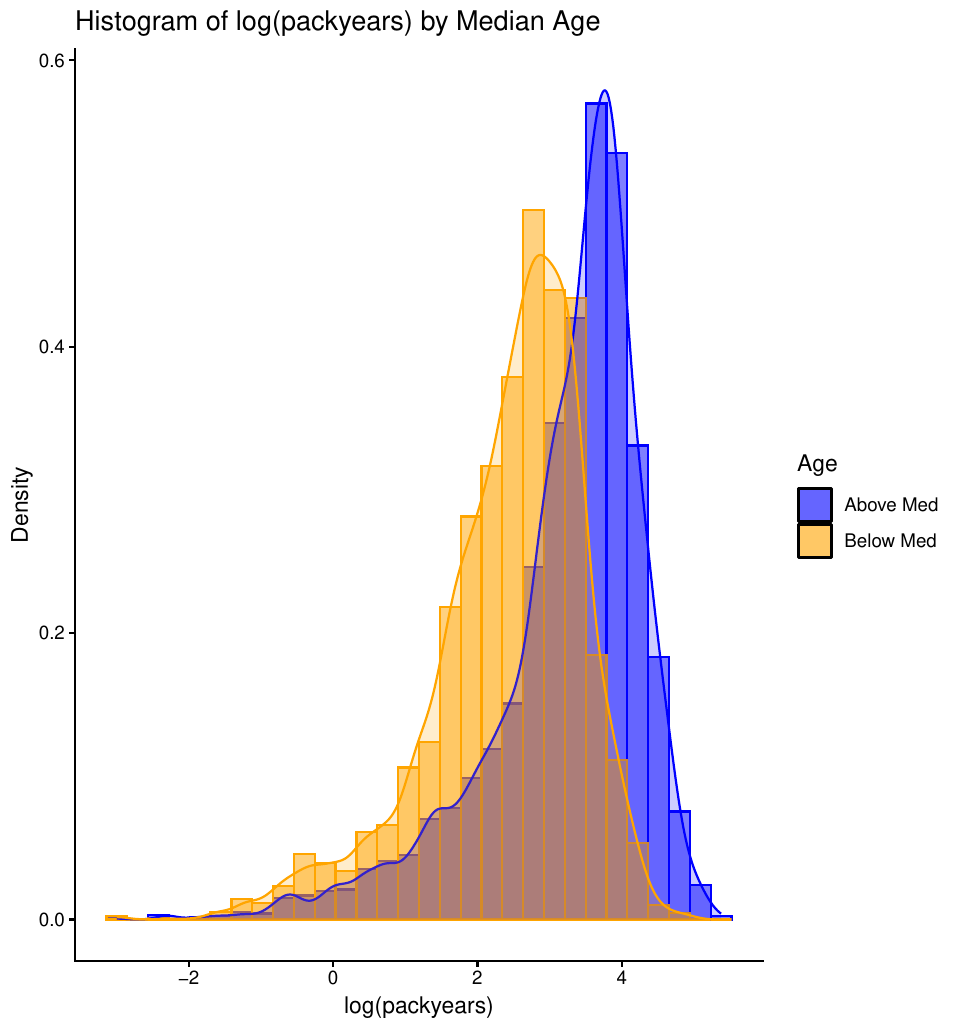}
\caption{The first plot is a histogram of the treatment for those younger than 48 (the median age) in 1987. The second plot is the same histogram, but for those older than 48. The combined plot is the third plot.}\label{fig:age_trt_hist}
\end{figure}

\section{Treatment Assignment Model Comparison}\label{sec:unif_comparison}

We consider a small simulation to investigate the treatment assignment distribution approach introduced in the main text Section 2.2.1 using a normal distribution centered at $1$. We compare that with a uniform treatment assignment, a kernel density treatment assignment with a na\"ive bandwidth (computed on the training data to match the training data distribution), and the correctly specified treatment assignment model. For each approach, the weights were capped to be at most 20 to ensure no prediction bands were of infinite length due to positivity violations in the simulations.

The setup is the same as the second scenario in Section 3.1 of the main text, except the treatment assignment distribution is $\mathcal{N}(1, .5^2)$.

\begin{table}[H]
\begin{center}
    \caption{Simulations: Data Generation}
    \label{tab:cov_gen_appendix}
    \begin{tabular}{|l|c|c|}
        \hline
        Variable & Generating Distribution \\
        \hline 
        $X_1$ & $\mathcal{N}(1, 1)$\\
        $X_2$ & $\mathcal{N}(1, 1)$\\
        $X_3$ & $\mathcal{N}(1, 4)$\\
        $T \mid \vect{X}$ & $\mathcal{N}(X_1 - X_2^2 + 0.5X_3, 20)$\\
        $T_{assign}$ & $\mathcal{N}(1, 0.5^2)$\\
        $Y(t) \mid \vect{X}$ & $\mathcal{N}(X_1 + 2X_2 + t + 5X_1^2, 9)$\\
         \hline
    \end{tabular}
\end{center}
\end{table}

For each simulation, 200 out of sample data points were generated. The kernel density weight bandwidth was computed using the default method in the \textit{density} function in R on the training data.  

All approaches use the oracle conditional mean, with the absolute difference non-conformity score, $|\mu(\vect{X}, T) - Y|$. The target coverage level was 90\%. Only treatments within one standard deviation (0.5) of the treatment of interest, $T = 1$, were considered. 

\begin{figure}[H]
    \centering
\includegraphics[scale = .6]{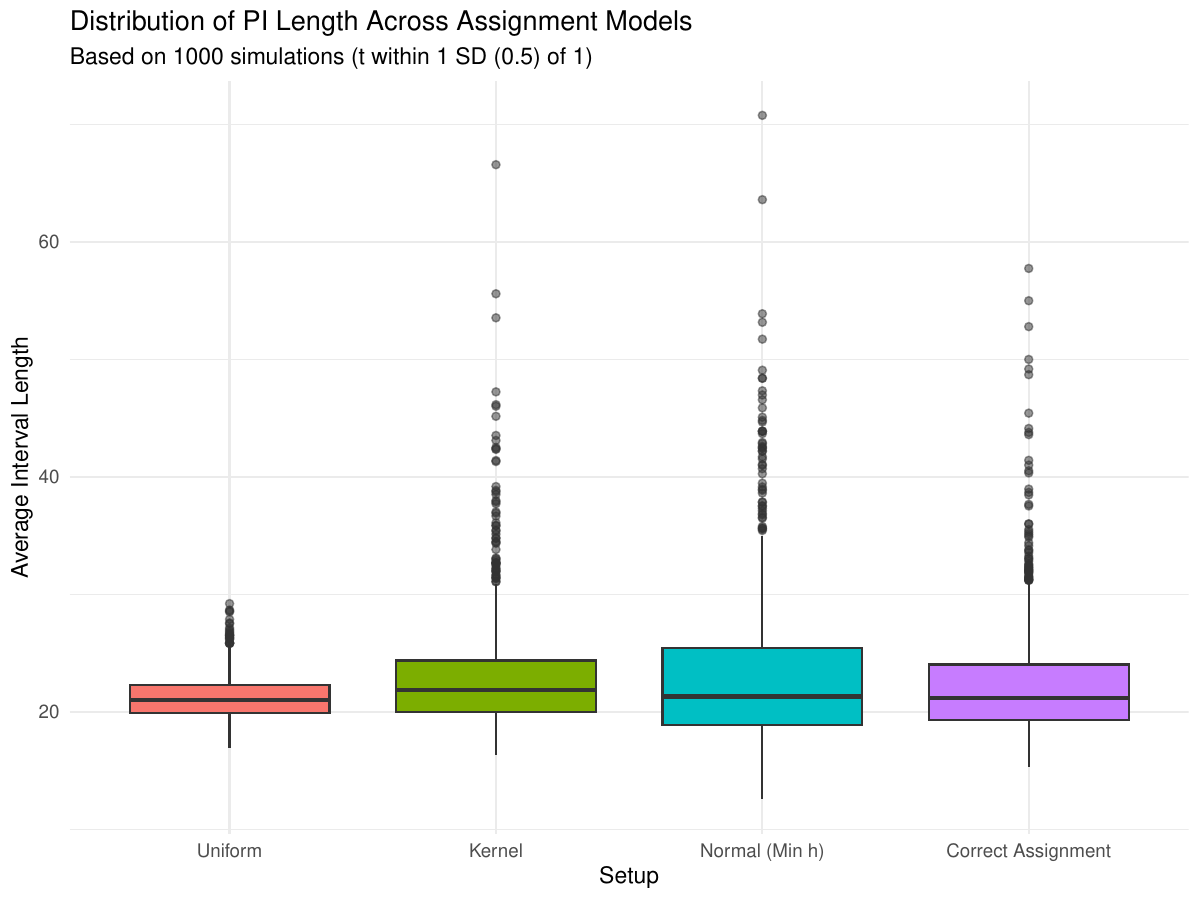}
\caption{The boxplots represent the distribution of the prediction interval lengths for all methods.}\label{fig:boxplot_comp_unif}
\end{figure}

\begin{figure}[H]
    \centering
\includegraphics[scale = .6]{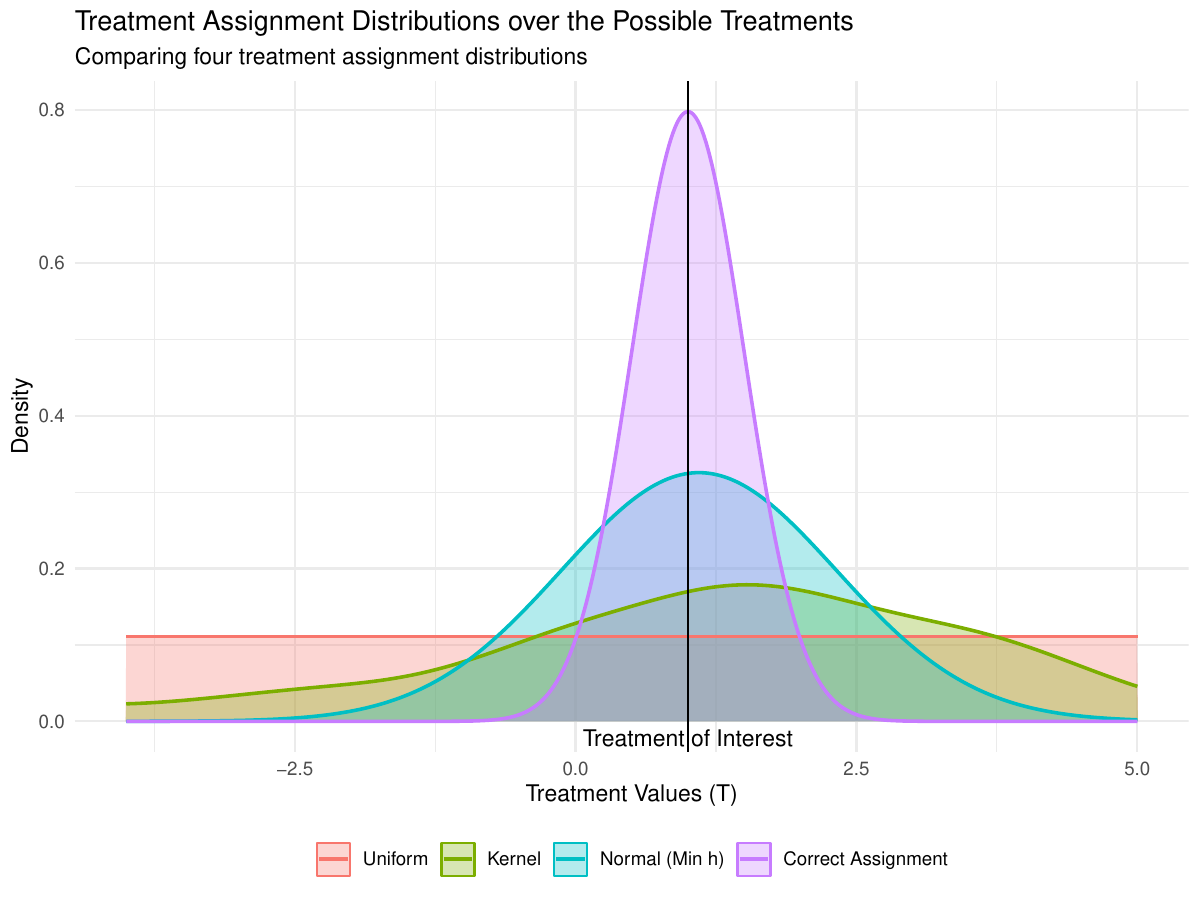}
\caption{The different treatment assignment distributions used from a randomly selected simulation.}\label{fig:dens_comparison}
\end{figure}

The coverage results can be found in~\cref{tab:sim1_results_unif_appendix}. A comparison of the distribution of the lengths can be found in~\cref{fig:boxplot_comp_unif}. An overlay of the different densities is given in~\cref{fig:dens_comparison}.

\begin{table}[H]
\centering
\caption{Simulation Result: Different Treatment Assignment Distributions} 
\label{tab:sim1_results_unif_appendix}
\begin{tabular}{lccc}
  \hline
Method & Coverage & Length \\ 
  \hline
Kernel & 0.907 (0.001) & 23.06 (0.15) \\ 
  Normal (Min h) & 0.896 (0.002) & 23.11 (0.20) \\ 
  Uniform & 0.901 (0.001) & 21.27 (0.06) \\ 
  Correct Assignment & 0.900 (0.001) & 22.47 (0.16) \\ 
   \hline
\end{tabular}
\label{tab:sim_results}
\end{table}

We can see the results for the methods are fairly similar. When capping weights, the uniform approach becomes very stabile compared to the other approaches. If the weights were not capped, the uniform approach has the most variable length, which is similar to ignoring the marginal treatment distribution when using the GPS as the denominator in a marginal structural model \citep{RobbinsHernan2000MSM}.

From the treatment assignment density plots in~\Cref{fig:dens_comparison} that, while not exact, conducting a search for the variance of a normal distribution as described in Section 2.2.1 places more of the density near to the treatment of interest compared to the kernel and uniform approaches. 

\section{Real Data Analysis Differing Deciles}

We conducted the real data analysis twice more, showing that if a practitioner wanted to put more weight on treatments in one decile without using a truncated treatment, they are able to.

\begin{figure}[H]
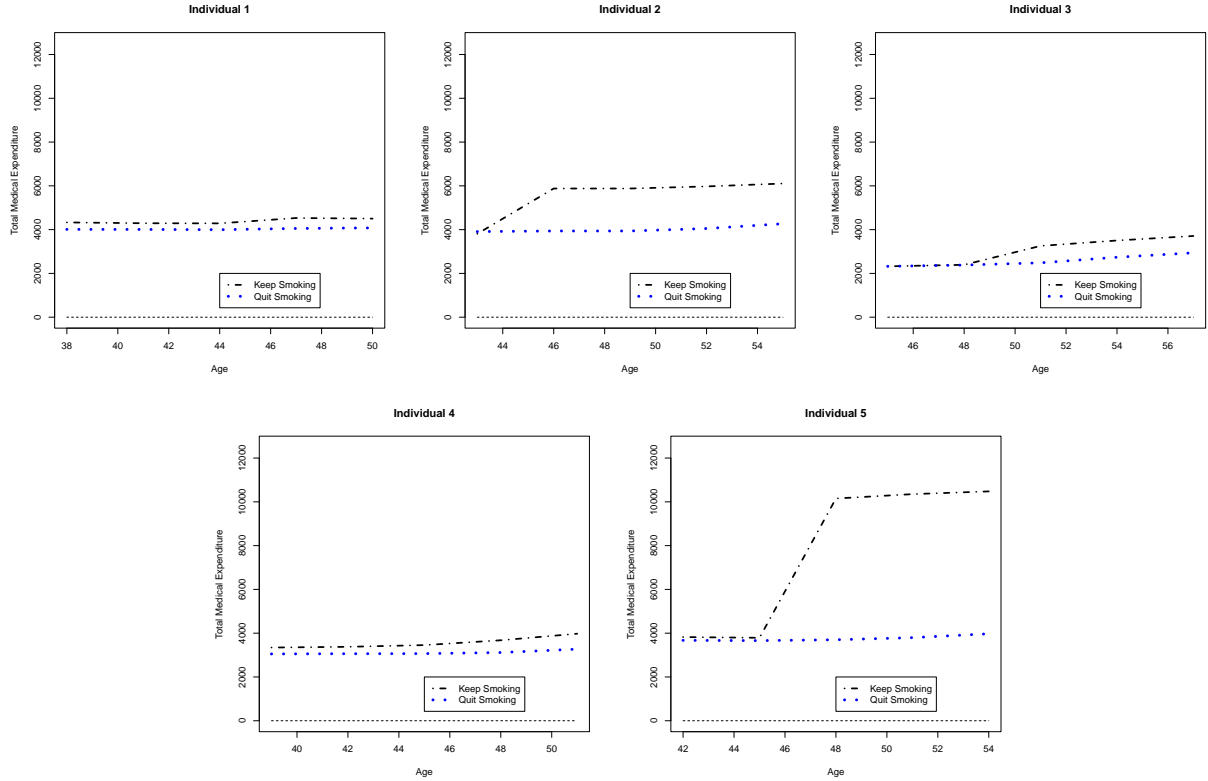

    \centering
\includegraphics[scale = .3]{images/ind1_12.pdf}
\includegraphics[scale = .3]{images/ind2_12.pdf}
\includegraphics[scale = .3]{images/ind3_12.pdf}
\includegraphics[scale = .3]{images/ind4_12.pdf}
\includegraphics[scale = .3]{images/ind5_12.pdf}

\caption{90\% upper conformal prediction intervals for continued smoking are given by the black dashed lines. 90\% upper prediction intervals for quitting smoking are given by the blue line. The treatment assignment distribution used a ratio of $k = 1/2$. }\label{fig:ind1_k1}
\end{figure}

\begin{figure}[H]
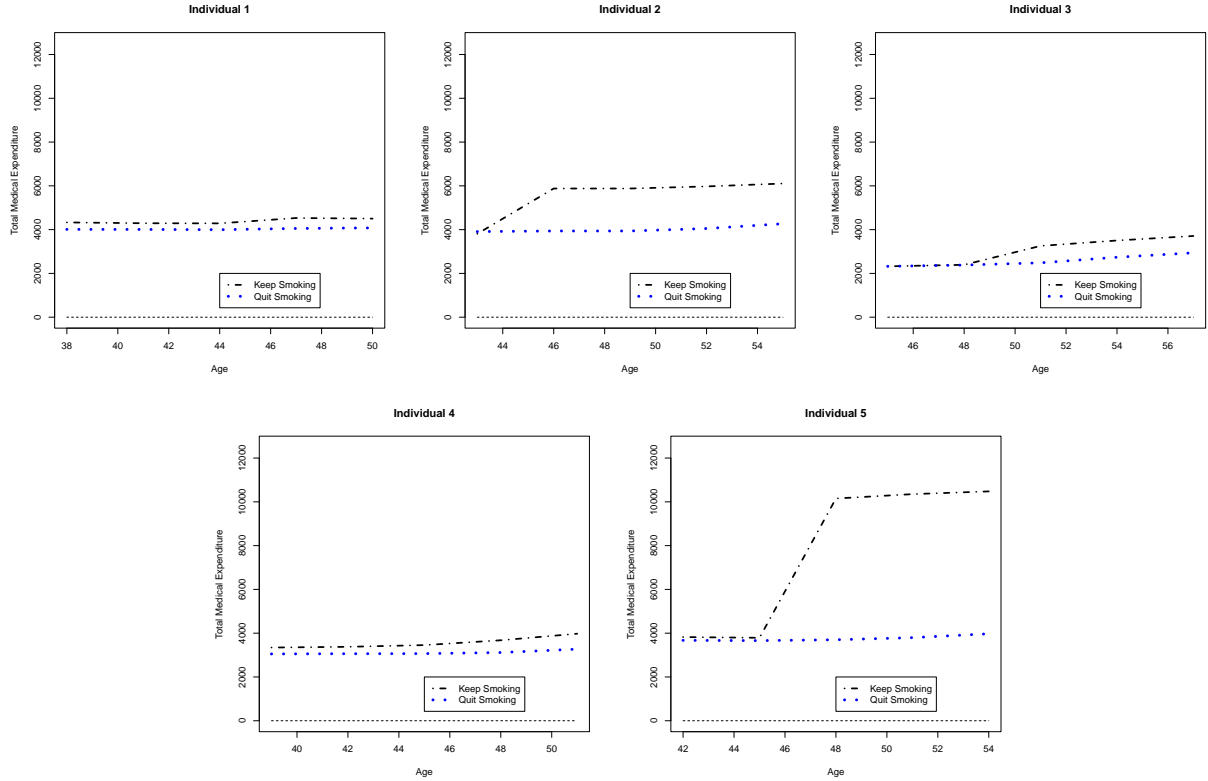

    \centering
\includegraphics[scale = .3]{images/ind1_14.pdf}
\includegraphics[scale = .3]{images/ind2_14.pdf}
\includegraphics[scale = .3]{images/ind3_14.pdf}
\includegraphics[scale = .3]{images/ind4_14.pdf}
\includegraphics[scale = .3]{images/ind5_14.pdf}

\caption{90\% upper conformal prediction intervals for continued smoking are given by the black dashed lines. 90\% upper prediction intervals for quitting smoking are given by the blue line. The last plot is a density plot of the packyear variable, along with the selected individual's packyear value plotted. The treatment assignment distribution used a ratio of $k = 1/4$. }
\label{fig:ind1_k2}
\end{figure}

If the treatment of interest and the observed treatment of an individual in the calibration set were in the same decile, the assignment was a Normal distribution with a mean equal to the mid-point of the treatment of interest's decile and a standard deviation equal to the standard deviation used to estimate the GPS. If the treatment of interest and the observed treatment of an individual in the calibration set were in different deciles, the assigned weight was $k$ the weight that would have been used if the observations were in the same decile, where $k$ is a practitioner selected parameter. This allowed us to use an approach similar to a kernel density, but with a larger weight for individuals in the calibration set with a similar packyear to the treatment of interest. Formulaically the weights used on the scores for a treatment of interest $t$ are, 
\[
W_i = \begin{cases}
\frac{\phi(T_i \mid d, s^2)}{\phi(T_i \mid \hat{\mu}(\vect{X}_i), s^2)}, & \text{if } T_i \text{ is in the same decile as } t, \\
\\
$k$ \times \frac{\phi(T_i \mid d, s^2)}{\phi(T_i \mid \hat{\mu}(\vect{X}_i), s^2)}, & \text{if } T_i \text{ is in a different decile than } t,
\end{cases}
\]
where $d$ is the mid-point of the decile containing $t$,
$\phi(\cdot\mid a, b)$ denotes the density of a Normal distribution with a mean of $a$ and a variance of $b$, $\hat{\mu}(\vect{x})$ denotes the estimated mean from the GPS, $s^2$ the mean squared error from the GPS model, and $k$ the weight for observations in the calibration set with a treatment outside of the same decile. The plots from the analysis redone with $k = 1/2$ can be found in~\cref{fig:ind1_k1} and for $k = 1/4$ in~\cref{fig:ind1_k2}.

We can see from the plots that adjusting $k$ does not change the results at all. Due to the weighted quantile, the final selected quantile is the same for $k = 1$, $1/2$, and $1/4$. This provides confidence that the patterns that emerge from using IPB are robust when the treatment allocation distributions are similar.

\section{Generalized Propensity Score} \label{sec:GPS_appendix}

The generalized propensity score was introduced by \cite{HiranoandImbens2004}, though the idea existed before then, see \cite{RobbinsHernan2000MSM} for one example. It is defined as the conditional density of $T$ given $\vect{X}$, $r(t, \vect{x}) = f(T = t\mid \vect{X} = \vect{x})$. A further generalization of the generalized propensity score is the propensity function (PF). Proposed by \cite{imaidyke2004}, they assume the conditional density of $T$ given $\vect{X}$ is indexed by $\vect{\psi}$. That is, $f(\cdot\mid \vect{X}) = f_{\vect{\psi}}(\cdot\mid \vect{X})$. If the density only depends on $\vect{X}$ through a finite dimensional quantity $\vect{\theta}_{\vect{\psi}}(\vect{X})$, then the propensity function can be rewritten as $f(\cdot\mid \vect{\theta}_{\vect{\psi}}(\vect{X}))$.

\cite{imaidyke2004} showed that, under strong ignorability, the PF has two useful properties. First, it is a balancing score, meaning that conditional on the PF, the treatment assignment is independent of the covariates. Second, given the PF, the treatment assignment and the potential outcomes are independent. Another way of phrasing that is, given the PF the treatment assignment is strongly ignorable, ($T \indep \mathcal{Y}\mid\vect{\theta}_{\vect{\psi}}(\vect{X})$). Practically, this means that high dimensional covariates that satisfy strong ignorability can be represented in a lower dimension without introducing bias. The PF still needs to be estimated using the data. It can be written as $\vect{\theta}(\vect{X}_i, \hat{\psi}) = \hat{\vect{\theta}}_i$, or $\hat{\theta}_i$ in the scalar case. 

\cite{RobbinsHernan2000MSM} introduced the idea of a marginal structural models. These are an extension of inverse probability treatment weighting to the continuous treatment setting with an assumed parametric form of the dose-response curve. For example, if the function is linear, $\mu(t) = \beta_0 + \beta_1 t$, the method estimates $\beta_0$ and $\beta_1$ by a least-squares calculation that minimizes
\[
\sum_i^N w_i(T_i)(Y_i - \beta_0 - \beta_1 T_i)^2,
\]
where $w_i(T_i)$ is a weight. A squared treatment term could also be included to be more robust. Both of these are examples of marginal structural models. A common weight is the inverse of the generalized propensity score, $w_i(T_i) = \frac{1}{f(T_i\mid \vect{X}_i)}$.

As in the binary case, these weights can be unstable or highly variable for observed treatment values that fall in the tails. \cite{RobbinsHernan2000MSM} recommend using stabilized weights $w_i(T)= \frac{f(T_i)}{f(T_i\mid \vect{X}_i)}$, where $f(T_i)$ is the marginal density of the treatment. This is similar to the idea of importance sampling. What might have occurred if the treatment was assigned at random instead of having been assigned according to the conditional density. That is, if it had been assigned according to $f(T_i)$ instead of $f(T_i\mid \vect{X}_i)$.

\cite{imaidyke2004} recommended subclassifying the data based on the PF, and estimating causal effects within each subclass. Assuming that the PF is correctly specified, within subclasses the covariates are evenly distributed across treatment levels. Then, within subclasses, the outcome can be modeled as a function of the treatment. For example, in each subclass model the response as $Y = \beta_0 + \beta_1 T + \beta_2 T^2 + \epsilon$. Then final estimates can be computed by weighting how many observations were in each subclass. For example, the average treatment effect at $T=t$ could be computed by $\hat{\mu}(t) = \sum\limits_j^g \frac{n_j}{n}\hat{f}_{j}(t)$, where $g$ is the number of subclass or groups, $n_j$ is the number of individuals in each subclass, $n$ is the total sample size, and $\hat{f}_i$ is the estimated response in the $i$-th subclass. 

\cite{HiranoandImbens2004} proposed using the GPS as an ``imputation'' tool. Their method consists of three steps. The first is to fit a parametric model to the treatment given the covariates. They specifically used the normal distribution for their conditional distribution, $T\mid X \sim \mathcal{N}(\vect{X}^T \beta, \sigma^2)$
\[
\hat{R}_i = \frac{1}{\sqrt{2\pi \hat{\sigma}^2}} \text{exp}(-\frac{1}{2\hat{\sigma}^2}(T_i - \vect{X}_i^T\hat{\beta})^2).
\]
The second step is to model the conditional expectation of $Y$ given $T$ and $\hat{R}$. They used a quadratic approximation,
\[
E(Y_i\mid T_i, \hat{R}_i) = \alpha_0 + \alpha_1 T_i + \alpha_2 T_i^2 + \alpha_3 \hat{R}_i + \alpha_4 \hat{R}_i^2 + \alpha_5 T_i \hat{R}_i,
\]
where the parameters, $\alpha_i$ are to be estimated. One common method is ordinary least squares. The final step is to estimate the average dose-response function at a particular treatment value, $t$. 
\[
\hat{\mu}(t) = \frac{1}{n} \sum_{i = 1}^n (\hat{\alpha}_0 + \hat{\alpha}_1 t + \hat{\alpha}_2 t^2 + \hat{\alpha}_3 \hat{r}(t, \vect{X}_i) + \hat{\alpha}_4 \hat{r}(t, \vect{X}_i)^2 + \hat{\alpha}_5 t \hat{r}(t, \vect{X}_i))
\]

This idea was extended by \cite{ZhaoVanDykImai2020}, but they used a smooth coefficient model that can vary as a function of the propensity score instead of a fixed parameter in each subclass. That is, they modeled the conditional response as $E(Y\mid T, \hat{\theta}) = f(\hat{\theta}) + g(\hat{\theta})T$, where $h(\cdot)$ and $g(\cdot)$ are assumed to be smooth and $\hat{\theta}$ is the estimated PF. 

The first step is to model the treatment given the covariates, and estimate a PF for each subject, $\hat{\theta}_i$. The next step is to fit a smooth coefficient model for the response given the treatment and the estimated PF, $E(Y\mid \hat{\theta}, T) = f(\hat{\theta}, T)$. This gives us an estimated function, $\hat{f}(\theta_i, T_i)$. The final step is to average over the $n$ estimated PF values we have to get an estimated average dose-response function at a particular treatment value of interest, $t$. $\hat{\mu}(t) = \frac{1}{n}\sum\limits_{i=1}^n \hat{f}(\hat{\theta}_i, t)$. The smooth coefficient model in the second step can be a flexible model, such as a generalized additive model (GAM). 

\cite{flores2012} extended the method of \cite{HiranoandImbens2004} by modeling the conditional expectation of $Y$ given $T$ and $\hat{R}$ non-parametrically. They then compute the estimated average dose-response at a given value of the treatment in the same way as \cite{HiranoandImbens2004} and \cite{ZhaoVanDykImai2020}.

\cite{flores2012} also proposed a second method that involves applying a Horvitz-Thompson weighting scheme to a continuous treatment. This can also be thought of as a generalized propensity score version of inverse probability weighting. 

To give the idea, assume that that a generalized propensity score has already been calculated, $\hat{R}_i$. First, bin the treatments. Then, let the number of bins increase while length of bins goes to zero. This mimics what happens when a treatment is truly continuous. Weight each response by 1 over the estimated GPS. That gives an estimate of the average dose-response at a specific value of $t$, 
\[
\hat{\mu}(t) =\frac{1}{n}\frac{\sum\limits_i^n \one(T_i \in \Delta)Y_i}{2h\hat{R}_i},
\]
where $h$ is a sequence of numbers going to 0 as $n \to \infty$ and $\Delta = [t - h, t + h]$. Note also that $P(T_i \in \Delta \approx 2h\hat{R}_i)$ for small values of $h$. 

This can be smoothed out by using a kernel function.
\[
\hat{\mu}(t) =\frac{\sum_i^n K_h(T_i - t)Y_iw_i(t)}{\sum_i^n K_h(T_i - t)w_i(t)},
\]
where $K_h(\cdot)$ is a kernel function, $h$ is a bandwidth satisfying $h \to 0$ and $nh \to \infty$ as $n \to \infty$, and $w_i(\cdot)$ are the inverse of the generalized propensity score.

Stabilized weights can also be used instead of unstabilized weights. This is recommended by \cite{ZhaoVanDykImai2020}. Notice that the weights now use the observed treatment instead of the treatment of interest. Let $\Tilde{K}(T_i - t) = \frac{f(T_i)}{f(T_i\mid \vect{X}_i)} K_h(T_i - t)$. Then our estimated average dose-response function is
\[
\hat{\mu}(t) =\frac{\sum_i^n \Tilde{K}_h(T_i - t)Y_i}{\sum_i^n \Tilde{K}_h(T_i - t)},
\]

Another approach that \cite{flores2012} looked at was to use a local linear regression.
\[
\hat{\mu}(t) = \frac{D_o(t)S_2(t) - D_1(t)S_1(t)}{S_0(t)S_2(t) - S_1^2(t)},
\]
where $S_j(t) = \sum_i^n \Tilde{K}_h(T_i - t)(T_i-t)^j$ and $D_j(t) = \sum_i^n \Tilde{K}_h(T_i - t)(T_i-t)^jY_i$.

	\section{BART} In general BART is designed to estimate models of the form
	\begin{equation}\label{6}
	Y = f(\vect{x}) + \epsilon,
	\end{equation} 
	\noindent where $\epsilon \sim N(0, \sigma^2)$, $Y$ is the response, and $\vect{x}$ is the vector of covariates \citep{actbart}. Because of how important the treatment group notation is in causal inference, it can be helpful to write \eqref{6} as 
	\begin{equation}\label{7}
	Y = f(t, \vect{x}) + \epsilon,
	\end{equation}
	where $\epsilon$ in \eqref{7} is the same as in \eqref{6}, $Y$ is the response, $t$ is the treatment assignment, and $\vect{x}$ is the vector of pretreatment variables. In the continuous treatment case, $f(t, \vect{x}) = E(Y(t)|\vect{X}=\vect{x}), \>\forall t \in \mathcal{T}$. 
    
\begin{figure}[H]
		\centering
		\includegraphics{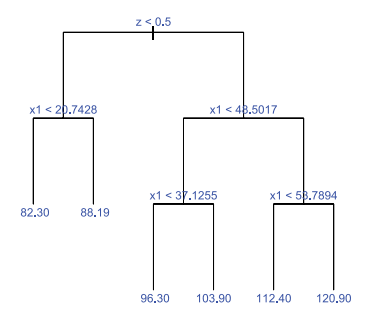}
		\caption{The binary tree fit to example data from \cite{bart}.}\label{fig1}
	\end{figure}

    Let $T$ denote a binary tree. The tree consists of all its decision rules that lead down to the bottom nodes.~\cref{fig1} shows an example of a single tree with one pretreatment variable and a treatment assignment variable. The decisions send the tree to the left, so if $Z = 0$ the tree would then check if $X_1 < 20.7428$ before ending at a node. Each bottom node has a parameter that is associated with it. For the $j^{th}$ node this parameter is $\mu_j$, which is the mean response of the observations that fall into the node. Let $\vect{M}$ denote the vector of mean values for $b$ nodes, $\vect{M} = \{\mu_1, ..., \mu_b\}$ where $b$ is the number of nodes for the tree. 
	
	Given a tree model $(T, \vect{M})$ and a pair of treatment assignments and pretreatment variables $(t, \vect{x})$, define $g(t, \vect{x}; T, \vect{M})$ as the value obtained by ``dropping'' $(t, \vect{x})$ down the tree $(T, \vect{M})$. For example, using the tree in~\cref{fig1} with $(z, x) = (1, 50)$, $g(1, 50, T, \vect{M}) = 112.40$. BART allows the user to approximate \eqref{7} with 
	\begin{equation*}\label{2}
		Y = g(z, \vect{x}; T_1, \vect{M_1}) + g(z, x; T_2, \vect{M_2})+ ... + g(z, \vect{x}; T_m, \vect{M_m}) + \epsilon,
	\end{equation*}
	where each $(T_j, \vect{M_j})$ denotes a single subtree model and $\epsilon$ is as before in \eqref{7}. The sum-of-trees model works to estimate this by fitting the first tree model, $g(t, \vect{x}, T_1, \vect{M_1})$. It then finds the residuals of the response and the first tree model, fits a tree to those residuals, and continues to do that a total of $m$ times. To avoid overfitting, a regularization prior is used on $(T_j, \vect{M_j})$. This allows each tree to only contribute a small part to the overall fit of the model. Each set of $(T_j, \vect{M_j})$ and $\sigma$ are treated as parameters in a model, they are given empirical priors, and they are redrawn during each MCMC iteration.
    
\begin{figure}[H]
		\centering
		\includegraphics{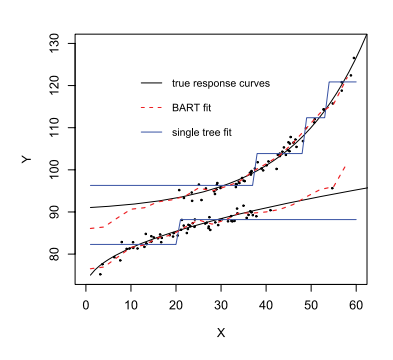}
		\caption{The BART fit using the same data as in Figure \ref{fig1}. The black points are the data, the dashed red line is the BART fit, the solid blue line is the fit with one tree, and the solid black line is the true curve \citep{bart}.}\label{fig2}
	\end{figure}

	During each MCMC iteration, each tree can shrink, grow, or change nonterminal node rules \citep{bart, actbart}. For more details on the algorithm and how it works see \cite{actbart}. The default prior has a preference for $T_j$ that have a small number of terminal nodes and a preference for $\mu_k$ that are $0$ (the response is centered in BART). These combined help prevent the model from overfitting. Of course, if the data suggests large trees, the prior for small trees and a small number of terminal nodes can be overcome. An example of a BART fit is given in~\cref{fig2}
	
\begin{figure}[H]
		\centering
		\includegraphics[scale = .5]{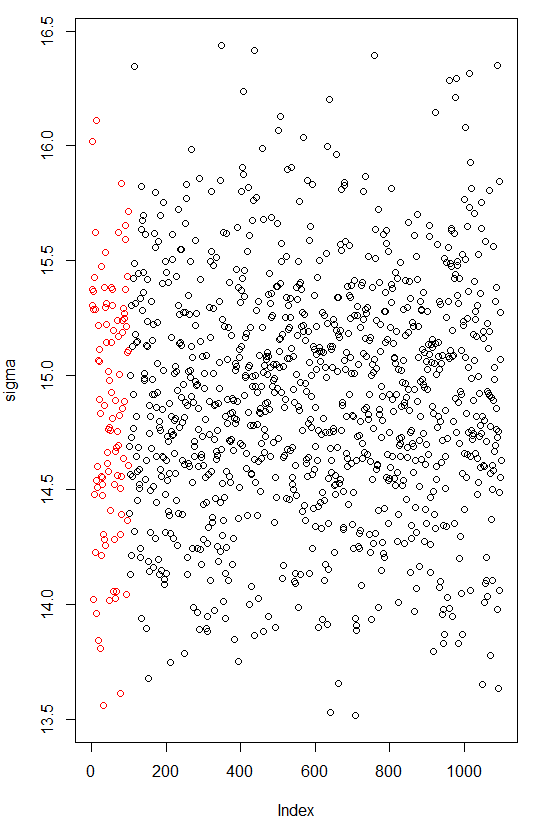}
		\caption{An example of BART convergence. The draws seen in red are the first 100 burn-in draws. These are not included in the final model. The true value of $\sigma$ was $15$. If the model did not converge, the draws of $\sigma$ would not be homogeneous.}\label{fig3}
	\end{figure}

	It was shown in \cite{actbart} that the performance of BART is competitive with the default values for the number of trees, the number of MCMC iterations, and the priors compared to model fitting methods that require cross-validation. The default number of trees is $200$, the default number of MCMC iterations is $1,000$ with a default of $100$ iterations for burn-in. This allows BART to be more ``automatic'' or easier to implement and apply compared to other methods. Because it is a sum-of-trees model, BART is able to capture linear terms easily. BART can also capture interaction terms. For example, a parent node could split at a binary treatment, $x_1$. Both nodes under that split could then split at another binary variable $x_2$. The terminal nodes under the splits at $x_2$ need not be monotonic, they could change based on the values that both $x_1$ and $x_2$ take. Because the tree can become larger or smaller during each iteration, it allows the level of interaction to be closely approximated \citep{bart}. As shown in \cite{actbart}, the algorithm tends to find very similar fits with different seeds. Convergence to the fit can be seen in~\cref{fig3} by looking at the draws of $\sigma$. The first one-hundred draws in red are burn in, these are not used for the final model.
	
    \subsection{Priors}
    We can write the estimation of $Y$ as 	
    \begin{equation*}\label{9}
	Y \approx \sum_{j = 1}^m g(x; T_j, M_j) + \epsilon,
    \end{equation*} 
    where $T_j$ denote the $j^{th}$ tree and $M_j$ denote the $j^{th}$ vector of terminal nodes, $M_j = (\mu_{1j}, \mu_{2j}, ..., \mu_{t_jj})$. 
    The parameters in the model are then $(T_j, M_j)$ and $\sigma$. 
    The priors are restricted to the form 
    \begin{equation*}\label{10}
    p(M_1, \ldots M_m, T_1, \ldots T_m) \propto\Big[\prod_{j = 1}^m p(M_j\mid T_j)p(T_j)\Big]p(\sigma),
    \end{equation*} 
    and
    \begin{equation*}
    p(M_j\mid T_j) = \prod_{i = 1} ^ {t_j} p(\mu_{ij}\mid T_j).
    \end{equation*}
    With this specification the pairs $(T_j, M_j)$ and $\sigma$ are a priori independent, making the calculation of the posteriors much easier \citep{actbart}.

    The prior for the trees has three parts:
    \begin{enumerate}
        \item The probability that a node is non-terminal is \(\frac{\alpha}{(1+d)^{\beta}}\), where $d$ is the depth of the node with $\alpha \in (0, 1)$ and $\beta \in [0, \infty)$ (The first node has depth 0, second 1, etc). The default values are $\alpha = 0.95$ and $\beta = 2$. 
        \item The probability that a variable is used at a split is uniform over the remaining predictors.
        \item The probability of the splitting value is uniform over the observed values of the variable at the split. A main advantage of this is that it is invariant to monotonic transformations of the response \citep{cart}. 
    \end{enumerate} 
    
    The prior used for $\sigma$ is an inverse chi-squared distribution with degrees of freedom $\nu = 3$ and scale $\lambda$ selected such that $P(\sigma < \hat{\sigma}) = 0.9$, where $\hat{\sigma}$ is the residual standard error from a least squares regression of $y$ on $(\vect{x}, t)$. The prior for $\mu_{ij} \mid T_j$  \(N(0, \sigma^2_\mu)\) \(\text{with }\sigma^2_{\mu} = \frac{0.5}{2 \sqrt{m}}\). It should be noted that BART transforms the response so that values range from -0.5 to 0.5 and are centered at zero. 

     The sampling conditional distribution is the product of the conditional distributions for the $j^{th}$ tree:
    \begin{equation*}\label{11}
    \theta_i = (T_i, M_i, \sigma, \vect{X}, t),
    \end{equation*} 
    \begin{equation*}\label{12}
    p(Y\mid\theta) = \prod_{k = 1}^{t_j} \prod_{i=1}^{n_k} f(y_{ki}\mid\theta_{k}),
    \end{equation*} 
    with
    \begin{equation*}\label{13}
    y_{k1}, ..., y_{kn_k}\mid\theta_k \sim N(\bar{\mu}_k, \sigma^2),\text{ } k = 1, ..., t_j \text{ for a fixed j},
    \end{equation*} 
    where $n_k$ represents the number of responses represented by the $k^{th}$ terminal node.
    The posterior is then 
    \begin{equation*}\label{14}
    p((T_1, M_1), ..., (T_m, M_m), \sigma \mid y)
    \end{equation*}

    To get samples from the posterior distribution, a Bayesian backfitting MCMC algorithm can be used. It involves $m$ successive draws from,
    \[
    (T_j, M_j)\mid T_{-j}, M_{-j}, \sigma, y, \quad j = 1,\ldots, m,
    \]
    where $T_{-j}, M_{-j}$ are the trees and parameters for all trees except the j-th tree. This is followed by a draw of $\sigma$ from the full conditional,
    \[
    \sigma \mid T_1, \ldots T_m, M_1,\ldots M_m, y.
    \]

    A set of $Q$ draws (after sufficient burnin) yields $f_1^*, \ldots f_Q^*$, which can be used as an approximate sample of size $Q$ from $p(f\mid y)$. This can be used to estimate the conditional mean,
    \[
    \frac{1}{Q}\sum_{q=1}^Q f_q^*(\vect{x}, t) \approx E(f(\vect{x}, t) \mid y).
    \]
    Similarly, conditional quantile estimates can be obtained by taking the quantiles of $f_q^*(\vect{x}, t)$. \citep{actbart, cart, hill2020bayesian} 

\end{document}